\documentclass[11pt]{article}

\usepackage[margin=25mm]{geometry}
\usepackage{setspace}
\usepackage[numbers]{natbib}
\usepackage{hyperref}
\usepackage{authblk}

\usepackage{amssymb}
\usepackage{pgf, interval}
    \intervalconfig{soft open fences}

\usepackage{graphicx}
    \graphicspath{{img/}}
\usepackage{booktabs}
    
\usepackage{caption}
\usepackage{standalone}
\usepackage{subcaption}

\usepackage{amssymb}

\usepackage{amsmath}
    \DeclareMathOperator*{\argmin}{arg\,min}

\usepackage{tikz}
    \usetikzlibrary{%
        arrows,
        backgrounds,
        decorations.pathreplacing,
        shapes.geometric,
        positioning,
    }

\definecolor{blue}{HTML}{0072B2}
\definecolor{green}{HTML}{009E73}
\definecolor{orange}{HTML}{D55E00}
\definecolor{pink}{HTML}{CC79A7}

\pgfdeclarelayer{background}
\pgfsetlayers{background,main}

\tikzstyle{every picture} += [remember picture]
\tikzstyle{na} = [baseline=-.5ex]

\tikzset{%
    queue/.pic={%
        code{%
            \node (rect) at (38.5mm, 10mm) {};
            \draw[thick] (0, 0) -- ++(40mm, 0) -- ++(0, 20mm) -- ++(-40mm, 0);
            \foreach \val in {0, ..., #1}{%
                \draw[thick] ([xshift=-\val*5mm] 40mm, 20mm) -- ++(0, -20mm);
            };

            \foreach \val/\lab/\size in {%
                0/1/\scriptsize,
                1/2/\scriptsize,
                3/c-1/\tiny,
                4/c/\scriptsize%
            }{%
                \node[draw, circle, thick, minimum size=9.5mm] (\lab)
                    at (55mm, 29mm - \val * 9.5mm) {\size$\lab$};
                \draw[-latex, thick] (rect.east) -- (\lab.west);
            };

            \node at (55mm, 11mm) {$\vdots$};
            \node at (5mm, 10mm) {$\cdots$};
        };
    },
    myarrow/.style={%
        line width=2mm,
        draw=gray!50,
        -triangle 60,
        postaction={draw=gray!50, line width=4mm, shorten >=6mm, -},
    },
    double -latex/.style args={#1 colored by #2 and #3}{%
        -latex,
        line width=#1,
        #2,
        postaction={%
            draw,
            -latex,
            #3,
            line width=(#1)/3,
            shorten <=(#1)/4,
            shorten >=4.5*(#1)/3
        },
    },
    mypointer/.style={%
        double -latex=1mm colored by gray!50 and gray!50,
    }
}

\newlength{\imgwidth}
\newlength{\tabwidth}
\title{%
    Segmentation analysis and the recovery of queuing parameters via the
    Wasserstein distance: a study of administrative data for patients with
    chronic obstructive pulmonary disease
}

\author[a]{%
    Henry Wilde \footnote{Corresponding author: \url{wildehd@cardiff.ac.uk}}%
}
\author[a]{Vincent Knight}
\author[a]{Jonathan Gillard}
\author[b]{Kendal Smith}

\affil[a]{School of Mathematics, Cardiff University}
\affil[b]{Cwm Taf Morgannwg University Health Board}

\date{}

\begin{document}
\maketitle%

\begin{abstract}
    This work uses a data-driven approach to analyse how the resource
    requirements of patients with chronic obstructive pulmonary disease (COPD)
    may change, quantifying how those changes impact the hospital system with
    which the patients interact. This approach is composed of a novel
    combination of often distinct modes of analysis: segmentation, operational
    queuing theory, and the recovery of parameters from incomplete data. By
    combining these methods as presented here, this work demonstrates that
    potential limitations around the availability of fine-grained data can be
    overcome. Thus, finding useful operational results despite using only
    administrative data. 

    The paper begins by finding a useful clustering of the population from this
    granular data that feeds into a multi-class \(M/M/c\) model, whose
    parameters are recovered from the data via parameterisation and the
    Wasserstein distance. This model is then used to conduct an informative
    analysis of the underlying queuing system and the needs of the population
    under study through several what-if scenarios.

    The analyses used to form and study this model consider, in effect, all
    types of patient arrivals and how those types impact the system. With that,
    this study finds that there are no quick solutions to reduce the impact of
    COPD patients on the system, including adding capacity to the system. In
    this analysis, the only effective intervention to reduce the strain caused
    by those presenting with COPD is to enact external policies which directly
    improve the overall health of the COPD population before they arrive at the
    hospital.
\end{abstract}

\section{Introduction}\label{sec:intro}

Population health research is increasingly based on data-driven methods (as
opposed to those designed solely by clinical experts) for patient-centred care
through the advent of accessible software and a relative abundance of electronic
data. However, many such methods rely heavily on detailed data — about both the
healthcare system and its population — which may limit research where
sophisticated data pipelines are not yet in place.

This work demonstrates a method of overcoming this, using routinely gathered,
administrative hospital data to build a clustering that feeds into a multi-class
queuing model, allowing for better understanding of the healthcare population
and the system with which they interact. Specifically, this work examines
records of patient spells from the National Health Service (NHS) Wales Cwm Taf
Morgannwg University Health Board (UHB) presenting chronic obstructive pulmonary
disease (COPD). COPD is a condition of particular interest to population health
research, and to Cwm Taf Morgannwg UHB, as it is known to often present as a
comorbidity in patients~\cite{Houben2019}, increasing the complexity of
treatments among those with the condition. Moreover, an internal report by NHS
Wales found the Cwm Taf Morgannwg UHB had the highest prevalence of the
condition across all the Welsh health boards.

This work draws upon several overlapping sources within mathematical research,
and this work contributes to the literature in three ways: to theoretical
queuing research by the estimation of missing queuing parameters with the
Wasserstein distance; to operational healthcare research through the weaving
together of the combination of methods used in this work despite data
constraints; and to public health research by adding to the growing body of
mathematical and operational work around a condition that is vital to understand
operationally, socially and medically. 

The remainder of the paper is structured as follows: Section~\ref{sec:intro}
provides a literature review, and an overview of the dataset and its clustering;
Section~\ref{sec:model} describes the queuing model used and the estimation of
its parameters; Section~\ref{sec:scenarios} presents several what-if scenarios
with insight provided by the model parameterisation and the clustering;
Section~\ref{sec:conclusion} concludes the paper. Although the data is
confidential and may not be published, a synthetic analogue has been
archived~\cite{Wilde2020synthetic} along with all the source code used in this
paper~\cite{Wilde2020github}.

\subsection{Literature review}\label{subsec:review}

Given the subject matter of this work, the relevant literature spans much of
operational research in healthcare, and the focus of this review is on the
critical topics of segmentation analysis, queuing models applied to hospital
systems, and the handling of missing or incomplete data for such queues.

\subsubsection{Segmentation analysis}

Segmentation analysis allows for the targeted analysis of otherwise
heterogeneous datasets and encompasses several techniques from operational
research, statistics and machine learning. One of the most desirable qualities
of this kind of analysis is the ability to glean and communicate simplified
summaries of patient needs to stakeholders within a healthcare
system~\cite{Vuik2016b, Yoon2020}. For instance, clinical profiling often forms
part of the broader analysis where each segment is summarised in a phrase or
infographic~\cite{Vuik2016a, Yan2019}.

The review for this work identified three commonplace groups of patient
characteristics used to segment a patient population: system utilisation
metrics; clinical attributes; and the pathway. The last is not used to segment
the patients directly, instead of grouping their movements through a healthcare
system, typically via process mining.~\cite{Arnolds2018}~and~\cite{Delias2015}
demonstrate how this technique can be used to improve the efficiency of a
hospital system as opposed to tackling the more relevant issue of
patient-centred care. The remaining characteristics can be segmented in a
variety of ways, but recent works tend to favour unsupervised methods —
typically latent class analysis (LCA) or clustering~\cite{Yan2018}.

LCA is a statistical, model-based method used to identify groups (called latent
classes) in data by relating its observations to some unobserved (latent),
categorical attribute. This attribute has multiple possible categories, each
corresponding to a latent class. The discovered relations enable the
observations to be separated into latent classes according to their maximum
likelihood class membership~\cite{Hagenaars2002,Lazarsfeld1968}. This method has
proved useful in the study of comorbidity patterns as
in~\cite{Kuwornu2014,Larsen2017} where combinations of demographic and clinical
attributes are related to various subgroups of chronic diseases.

Similarly to LCA, clustering identifies groups (clusters) in data to produce
labels for its instances. However, clustering includes a wide variety of methods
where the common theme is to maximise homogeneity within, and heterogeneity
between, each cluster~\cite{Everitt2011}. The \(k\)-means paradigm is the most
popular form of clustering in literature. The method iteratively partitions
numerical data into \(k \in \mathbb N\) distinct parts where \(k\) is fixed a
priori. This method has proved popular as it is easily scalable, and its
implementations are concise~\cite{Olafsson2008,Wu2009}. In addition to
\(k\)-means, hierarchical clustering methods can be useful if a suitable number
of parts cannot be found initially~\cite{Vuik2016a}. However, supervised
hierarchical segmentation methods such as classification and regression trees
(as in~\cite{Harper2006}) have been used where an existing, well-defined, label
is of particular significance.

\subsubsection{Queuing models}

Since the seminal works by Erlang~\cite{Erlang1917,Erlang1920} established the
core concepts of queuing theory, the application of queues and queuing networks
to real services has become abundant, including the healthcare service. By
applying these models to healthcare settings, many aspects of the underlying
system can be studied. A common area of study in healthcare settings is of
service capacity.~\cite{McClain1976} is an early example of such work where
acute bed capacity was determined using hospital occupancy data. Meanwhile, more
modern works such as~\cite{Palvannan2012,Pinto2014} consider more extensive
sources of data to build their queuing models.  Moreover, the output of a model
is catered more towards being actionable --- as is the prerogative of
operational research. For instance,~\cite{Pinto2014} devises new categorisations
for both hospital beds and arrivals that are informed by the queuing model. A
further example is~\cite{Komashie2015} where queuing models are used to measure
and understand satisfaction among patients and staff.

In addition to these theoretic models, healthcare queuing research has expanded
to include computer simulation models. The simulation of queues, or networks
thereof, have the benefit of adeptly capturing the stochastic nuances of
hospital systems over their theoretic counterparts. Example areas include the
construction and simulation of Markov processes via process
mining~\cite{Arnolds2018,Rebuge2012}, and patient flow~\cite{Bhattacharjee2014}.
Regardless of the advantages of simulation models, a prerequisite is reliable
software with which to construct those simulations. A common approach to
building simulation models of queues is to use a graphical user interface such
as Simul8. These tools have the benefits of being highly visual, making them
attractive to organisations looking to implement queuing models without
necessary technical expertise, including the NHS.~\cite{Brailsford2013}
discusses the issues around operational research and simulation being taken up
in the NHS despite the availability of intuitive software packages like Simul8.
However, they do not address a core principle of good simulation work:
reproducibility. The ability to reliably reproduce a set of results is of great
importance to scientific research but remains an issue in simulation research
generally~\cite{Fitzpatrick2019}. When considering issues with reproducibility
in scientific computing (simulation included), the source of any concerns is
often with the software used~\cite{Ivie2018}. Using well-developed, open-source
software can alleviate issues around reproducibility and reliability as how they
are used involve less uncertainty and require more rigour than ‘drag-and-drop’
software. One example of such a piece of software is Ciw~\cite{Palmer2019}. Ciw
is a discrete event simulation library written in Python that is fully
documented and tested. The simulations constructed and studied in
Sections~\ref{sec:model}~and~\ref{sec:scenarios} utilise this library and aid
the overall reproducibility of this work.

\subsubsection{Handling incomplete queue data}

As is discussed in other parts of this section, the data available in this work
is not as detailed as in other comparative works. Without access to such data
--- but intending to gain insight from what is available --- it is
imperative to bridge the gap left by the incomplete data.

Moreover, it is often the case that in practical situations where suitable data
is not (immediately) available, further inquiry in that line of research will
stop. Queuing models in healthcare settings appear to be such a case; the line
ends at incomplete queue data.~\cite{Asanjarani2017} is a bibliographic work
that collates articles on the estimation of queuing system characteristics ---
including their parameters. Despite its breadth of almost 300 publications from
1955, only two articles have been identified as being applied to
healthcare:~\cite{Mohammadi2012,Yom2014}. Both works are concerned with
customers who can re-enter services during their time in the queuing system,
which is mainly of value when considering the effect of unpredictable behaviour
in intensive care units, for instance.~\cite{Mohammadi2012} seeks to approximate
service and re-service densities through a Bayesian approach and by filtering
out those customers seeking to be serviced again. On the other
hand,~\cite{Yom2014} considers an extension to the \(M/M/c\) queue with direct
re-entries. The devised model is then used to determine resource requirements in
two healthcare settings.

Aside from healthcare-specific works, the approximation of queue parameters has
formed a part of relevant modern queuing research. However, the scope is
primarily focused on theoretic approximations rather than by
simulation.~\cite{Djabali2018,Goldenshluger2016} are two such recent works that
consider an underlying process to estimate a general service time distribution
in single server and infinite server queues respectively.

\subsection{Overview of the dataset and its clustering}\label{subsec:overview}

The Cwm Taf Morgannwg UHB provided the dataset used in this work. The
dataset contains an administrative summary of 5,231 patients presenting COPD
from February 2011 through March 2019 totalling 10,861 spells. A patient
(hospital) spell is defined as the continuous stay of a patient using a hospital
bed on premises controlled by a healthcare provider and is made up of one or
more patient episodes~\cite{NHS2020}. The following attributes describe the
spells included in the dataset:
\begin{itemize}
    \item Personal identifiers and information, i.e.\ patient and spell ID
        numbers, and identified gender;
    \item Admission/discharge dates and approximate times;
    \item Attributes summarising the clinical path of the spell including
        admission/discharge methods, and the number of episodes, consultants and
        wards in the spell;
    \item International Classification of Diseases (ICD) codes and primary
        Healthcare Resource Group (HRG) codes from each episode;
    \item Indicators for any COPD intervention. The value for any given instance
        in the dataset (i.e. a spell) is one of no intervention, pulmonary
        rehabilitation (PR), specialist nursing (SN), and both interventions;
    \item Charlson Comorbidity Index (CCI) contributions from several long term
        conditions (LTCs) as well as indicators for some other conditions such
        as sepsis and obesity. CCI is useful in anticipating hospital
        utilisation as a measure for the burdens associated with
        comorbidity~\cite{Simon2011};
    \item Rank under the 2019 Welsh Index of Multiple Deprivation (WIMD),
        indicating relative deprivation of the postcode area the patient lives
        in which is known to be linked to COPD prevalence and
        severity~\cite{Collins2018,Sexton2016,Steiner2017}.
\end{itemize}

In addition to the above, the following attributes were engineered for each
spell:
\begin{itemize}
    \item Age and spell cost data were linked to approximately half of the
        spells in the dataset from another administrative dataset provided by
        the Cwm Taf Morgannwg UHB;
    \item The presenting ICD codes were generalised to their categories
        according to NHS documentation and counts for each category were
        attached. This reduced the number of values from
        1,926 codes to 21 categories;
    \item A measure of admission frequency was calculated by taking the number
        of COPD-related admissions in the last twelve months linked to the
        associated patient ID number.
\end{itemize}

Although there is a fair amount of information here, it is limited to
COPD-related admissions. Therefore, rather than segmenting the patients
themselves, the spells will be. The clustering algorithm of choice is a variant
of \(k\)-means, called \(k\)-prototypes, allows for the clustering of mixed-type
data by performing \(k\)- means on the numeric attributes and \(k\)-modes on the
categoric. Both \(k\)-prototypes and \(k\)-modes were presented
in~\cite{Huang1998}.

The attributes included in the clustering encompass both utilisation metrics and
clinical attributes relating to the spell. They comprise the summative clinical
path attributes, the CCI contributions and condition indicators, the WIMD rank,
length of stay (LOS), COPD intervention status, and the engineered attributes
(not including age and costs due to lack of coverage).

To determine the optimal number of clusters, \(k\), the knee point detection
algorithm introduced in~\cite{Satopaa2011} was used with a range of potential
values for \(k\) from two to 10. This range was chosen based on what may be
considered feasibly informative to stakeholders. The knee point detection
algorithm can be considered a deterministic version of the widely known `elbow
method' for determining the number of clusters. Applying this algorithm
revealed an optimal value for \(k\) of four, but both three and five clusters
were considered. Both of these cases were eliminated due to a lack of clear
separation in the characteristics of the clusters. Additionally, the
initialisation method used for \(k\)-prototypes was presented
in~\cite{Wilde2020} as it was found to give an improvement in the clustering
over other initialisation methods.

\begin{table}
    \centering
    \resizebox{\tabwidth}{!}{%
        \begin{tabular}{llrrrrr}
        \toprule
               &        &  Cluster &          &          &           & Population \\
               &        &        0 &        1 &        2 &         3 &            \\
        \midrule
        \textbf{Characteristics} & \textbf{Percentage of spells} &     9.91 &    19.27 &    69.39 &      1.44 &     100.00 \\
               & \textbf{Mean spell cost, £} &  8051.23 &  2309.63 &  1508.41 &  17888.43 &    2265.40 \\
               & \textbf{Percentage of recorded costs} &    29.01 &    19.38 &    48.20 &      3.40 &     100.00 \\
               & \textbf{Median age} &    77.00 &    77.00 &    71.00 &     82.00 &      73.00 \\
               & \textbf{Minimum LOS} &    12.82 &    -0.00 &    -0.02 &     48.82 &      -0.02 \\
               & \textbf{Mean LOS} &    25.30 &     6.46 &     4.11 &     75.36 &       7.68 \\
               & \textbf{Maximum LOS} &    51.36 &    30.86 &    16.94 &    224.93 &     224.93 \\
               & \textbf{Median COPD adm. in last year} &     2.00 &     1.00 &     1.00 &      2.00 &       1.00 \\
               & \textbf{Median no. of LTCs} &     2.00 &     3.00 &     1.00 &      3.00 &       1.00 \\
               & \textbf{Median no. of ICDs} &     9.00 &     8.00 &     5.00 &     11.00 &       6.00 \\
               & \textbf{Median CCI} &     9.00 &    20.00 &     4.00 &     18.00 &       4.00 \\
        \textbf{Intervention prevalence} & \textbf{None, \%} &    80.20 &    83.42 &    65.76 &     89.74 &      70.94 \\
               & \textbf{PR, \%} &    15.80 &    13.43 &    27.97 &      8.97 &      23.69 \\
               & \textbf{SN, \%} &     3.81 &     2.87 &     4.63 &      1.28 &       4.16 \\
               & \textbf{Both, \%} &     0.19 &     0.29 &     1.63 &      0.00 &       1.21 \\
        \textbf{LTC prevalence} & \textbf{Pulmonary disease, \%} &   100.00 &   100.00 &   100.00 &    100.00 &     100.00 \\
               & \textbf{Diabetes, \%} &    19.05 &    28.14 &    14.84 &     25.00 &      17.96 \\
               & \textbf{AMI, \%} &    13.85 &    22.93 &     8.76 &     16.03 &      12.10 \\
               & \textbf{CHF, \%} &    12.45 &    53.85 &     0.00 &     26.28 &      11.99 \\
               & \textbf{Renal disease, \%} &     7.53 &    19.54 &     1.92 &     17.95 &       6.10 \\
               & \textbf{Cancer, \%} &     7.62 &    12.23 &     2.93 &     10.90 &       5.30 \\
               & \textbf{Dementia, \%} &     6.88 &    21.26 &     0.00 &     26.92 &       5.17 \\
               & \textbf{CVA, \%} &     8.64 &    13.33 &     0.70 &     19.87 &       4.20 \\
               & \textbf{PVD, \%} &     4.37 &     7.69 &     2.27 &      5.77 &       3.57 \\
               & \textbf{CTD, \%} &     5.11 &     4.25 &     3.11 &      4.49 &       3.54 \\
               & \textbf{Obesity, \%} &     2.51 &     3.01 &     1.49 &      7.69 &       1.97 \\
               & \textbf{Metastatic cancer, \%} &     1.58 &     4.49 &     0.00 &      0.64 &       1.03 \\
               & \textbf{Paraplegia, \%} &     1.30 &     3.73 &     0.24 &      0.64 &       1.02 \\
               & \textbf{Diabetic compl., \%} &     0.19 &     0.86 &     0.48 &      1.92 &       0.54 \\
               & \textbf{Peptic ulcer, \%} &     1.58 &     0.81 &     0.23 &      1.28 &       0.49 \\
               & \textbf{Sepsis, \%} &     1.77 &     0.91 &     0.15 &      1.92 &       0.48 \\
               & \textbf{Liver disease, \%} &     0.28 &     0.48 &     0.23 &      0.00 &       0.28 \\
               & \textbf{C. diff, \%} &     0.74 &     0.10 &     0.01 &      0.64 &       0.11 \\
               & \textbf{Severe liver disease, \%} &     0.19 &     0.43 &     0.00 &      0.00 &       0.10 \\
               & \textbf{MRSA, \%} &     0.28 &     0.05 &     0.03 &      1.28 &       0.07 \\
               & \textbf{HIV, \%} &     0.00 &     0.00 &     0.03 &      0.00 &       0.02 \\
        \bottomrule
        \end{tabular}
    }\caption{%
        A summary of clinical and condition-specific characteristics for each
        cluster and the population. A negative length of stay indicates that the
        patient died prior to arriving at the hospital.
    }\label{tab:summary}
\end{table}

A summary of the spells is provided in Table~\ref{tab:summary}. This table
separates each cluster and the overall dataset (referred to as the population).
From this table, helpful insights can be gained about the segments identified by
the clustering. For instance, the needs of the spells in each cluster can be
summarised succinctly:
\begin{itemize}
    \item Cluster 0 represents those spells with relatively low clinical
        complexity but high resource requirements. The mean spell cost is almost
        four times the population average, and the shortest spell is almost two
        weeks long. Moreover, the median number of COPD-related admissions in
        the last year is elevated, indicating that patients presenting in this
        way require more interactions with the system.
    \item Cluster 1, the second-largest segment, represents the spells with
        complex clinical profiles despite lower resource requirements.
        Specifically, the spells in this cluster have the highest median CCI and
        number of LTCs, and the highest condition prevalence across all clusters
        but the second-lowest length of stay and spell costs.
    \item Cluster 2 represents the majority of spells and those where resource
        requirements and clinical complexities are minimal; these spells have
        the shortest lengths, and the patients present with fewer diagnoses and
        a lower median CCI than any other cluster. In addition to this, the
        spells in Cluster 2 have the highest intervention prevalence. However,
        they have the lowest condition prevalence across all clusters.
    \item Cluster 3 represents the smallest section of the population but
        perhaps the most critical: spells with high complexity and high resource
        needs. The patients within Cluster 3 are the oldest in the population
        and are some of the most frequently returning despite having the lowest
        intervention rates. The lengths of stay vary between seven and 32 weeks,
        and the mean spell cost is almost eight times the population average.
        This cluster also has the second-highest median CCI, and the highest
        median number of concurrent diagnoses.
\end{itemize}

\begin{figure}
    \centering
    \begin{subfigure}{.5\imgwidth}
        \includegraphics[width=\linewidth]{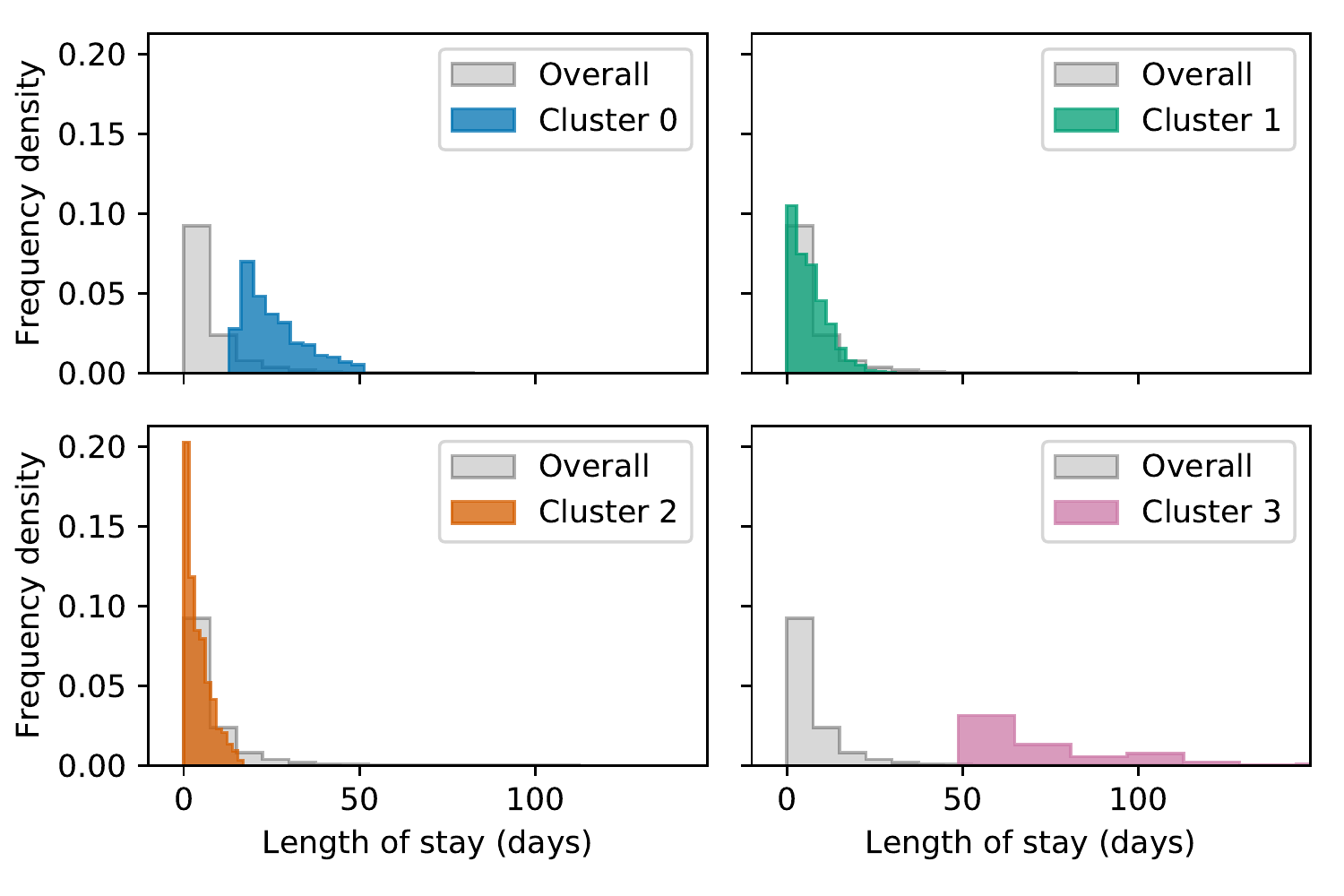}
        \caption{}\label{fig:cluster_los}
    \end{subfigure}\hfill%
    \begin{subfigure}{.5\imgwidth}
        \includegraphics[width=\linewidth]{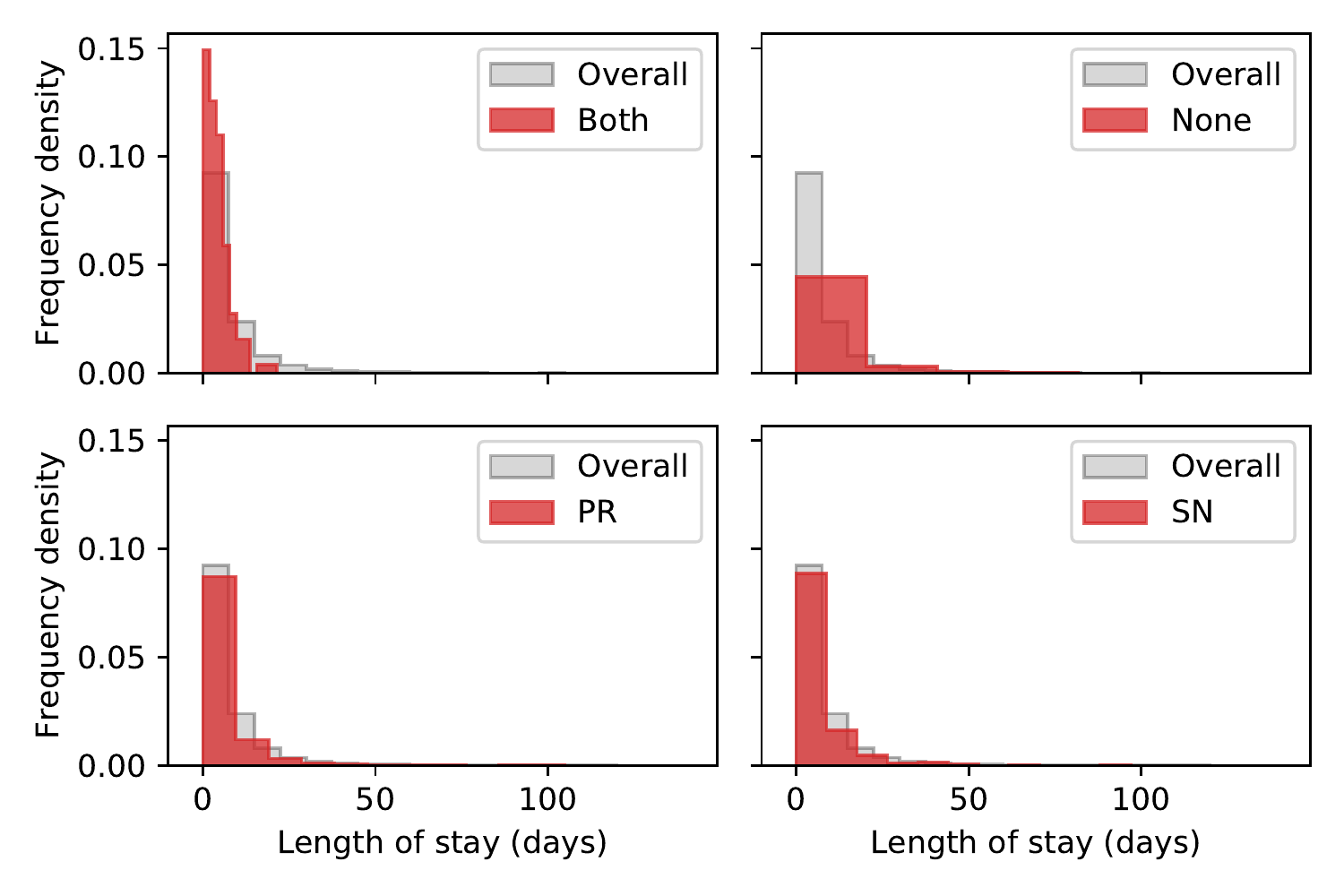}
        \caption{}\label{fig:intervention_los}
    \end{subfigure}
    \caption{%
        Histograms for length of stay by (\subref{fig:cluster_los}) cluster and
        (\subref{fig:intervention_los}) intervention.
    }\label{fig:los}
\end{figure}

\begin{figure}
    \centering
    \begin{subfigure}{.5\imgwidth}
        \includegraphics[width=\linewidth]{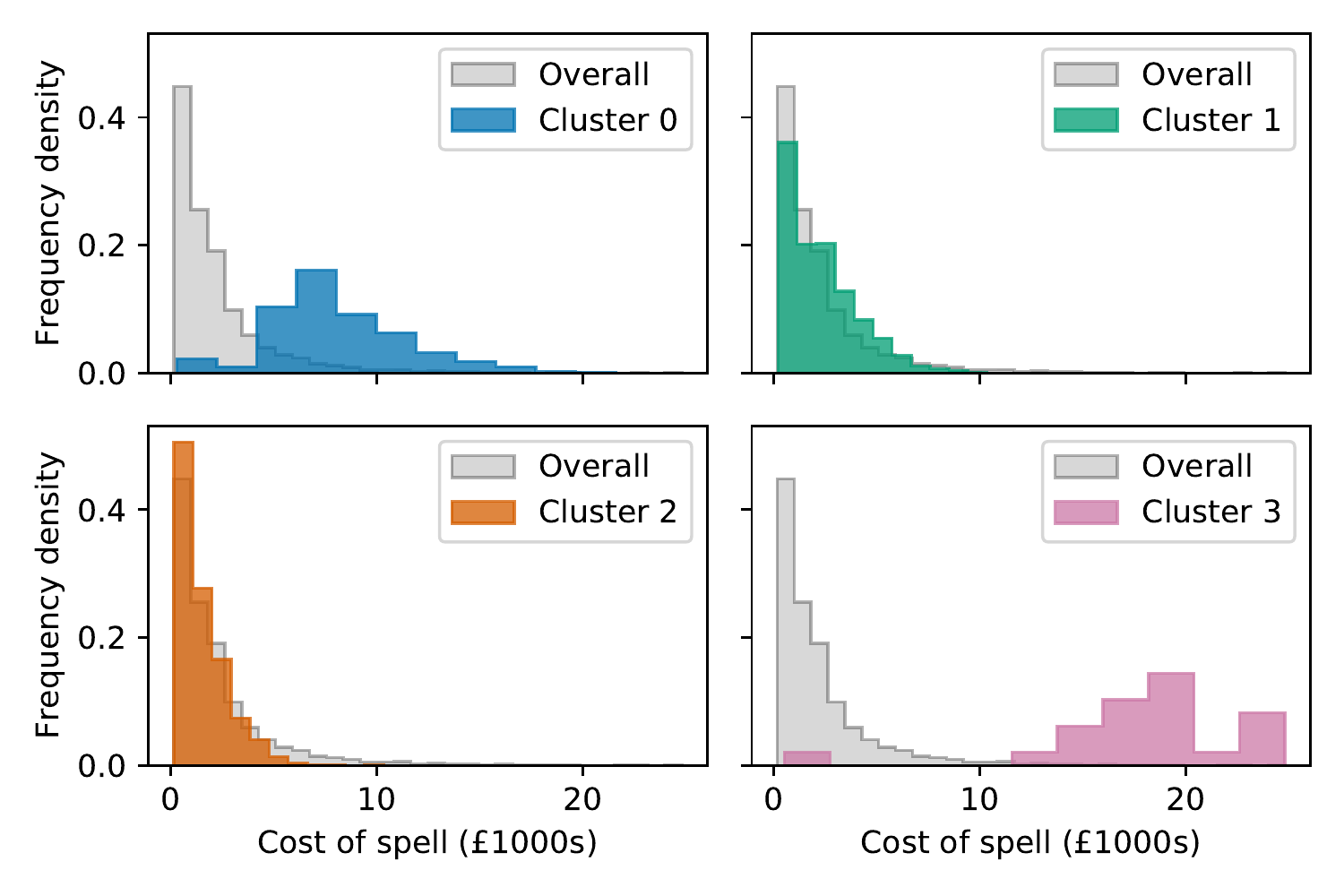}
        \caption{}\label{fig:cluster_cost}
    \end{subfigure}\hfill%
    \begin{subfigure}{.5\imgwidth}
        \includegraphics[width=\linewidth]{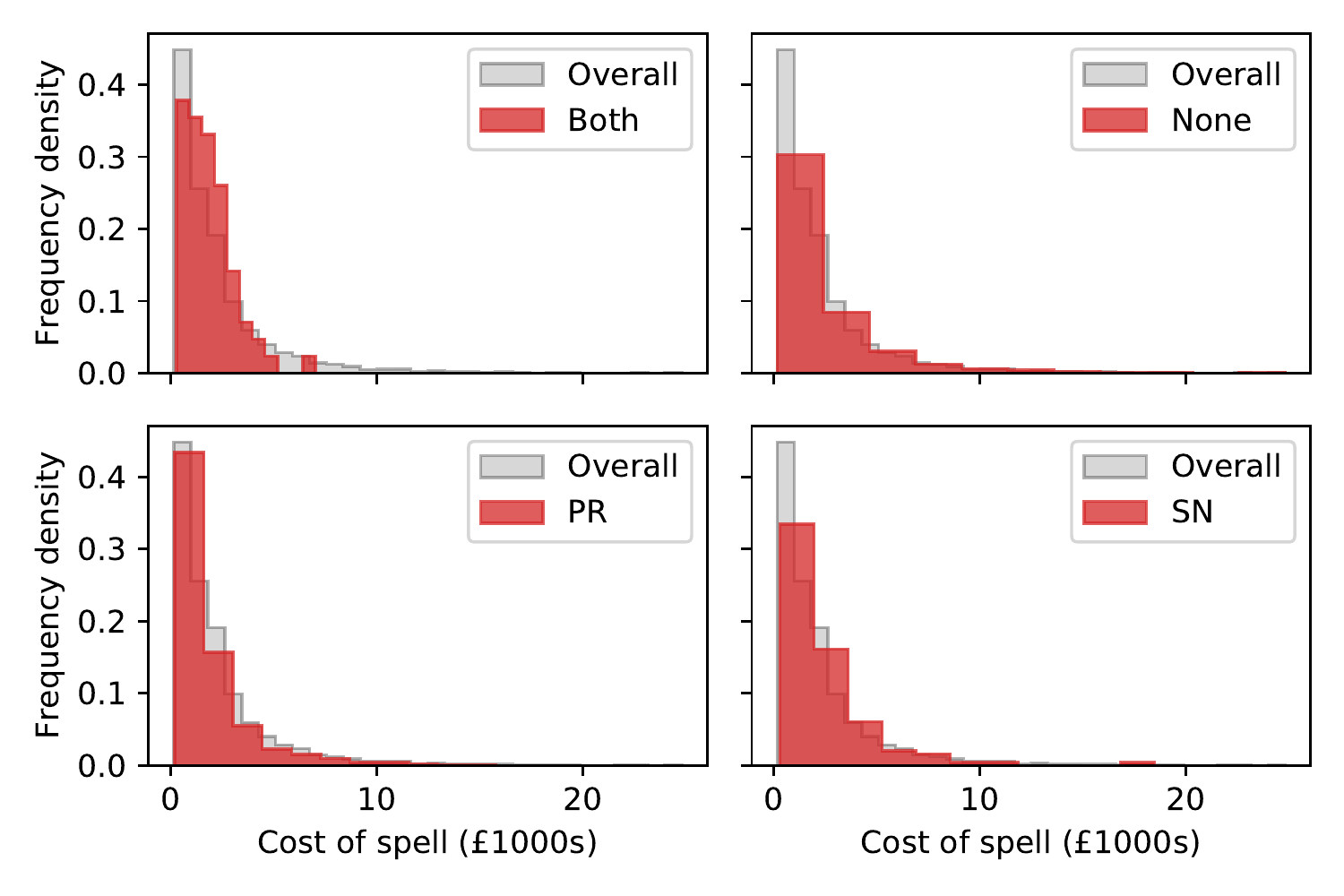}
        \caption{}\label{fig:intervention_cost}
    \end{subfigure}
    \caption{%
        Histograms for spell cost by (\subref{fig:cluster_cost}) cluster and
        (\subref{fig:intervention_cost}) intervention.
    }\label{fig:cost}
\end{figure}

\begin{figure}
    \centering
    \begin{subfigure}{.5\imgwidth}
        \includegraphics[width=\linewidth]{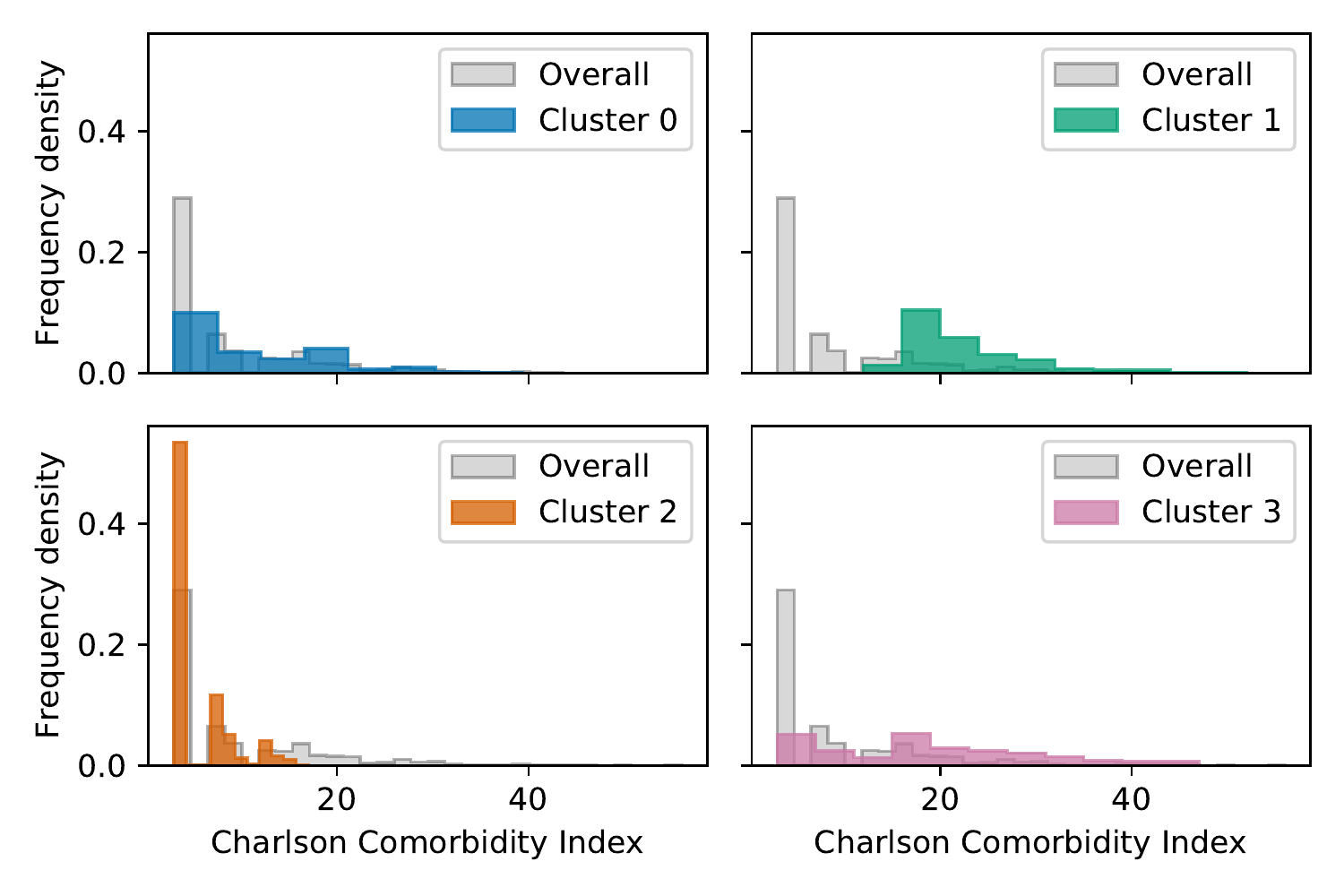}
        \caption{}\label{fig:cluster_charlson}
    \end{subfigure}\hfill%
    \begin{subfigure}{.5\imgwidth}
        \includegraphics[width=\linewidth]{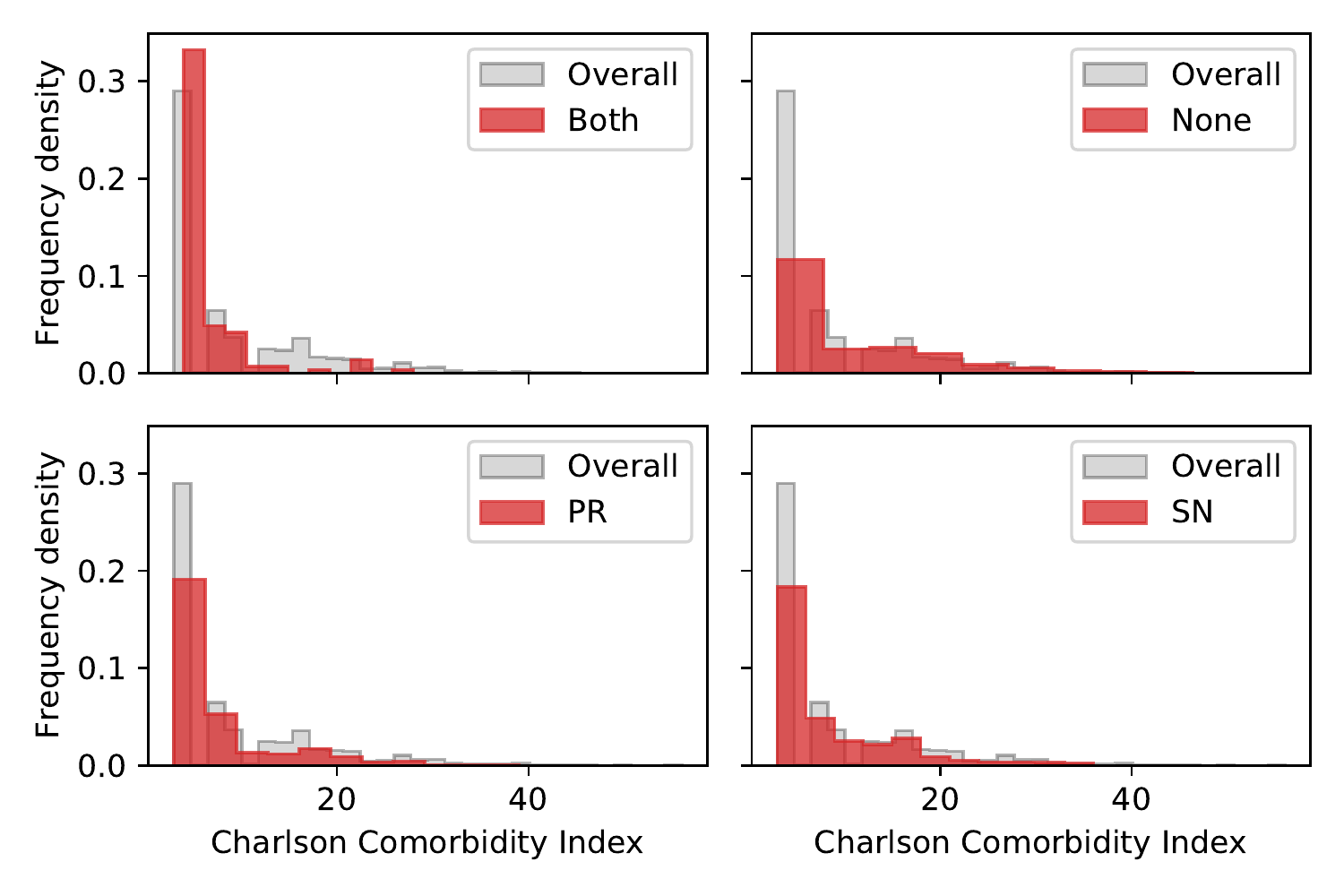}
        \caption{}\label{fig:intervention_charlson}
    \end{subfigure}
    \caption{%
        Histograms for CCI by (\subref{fig:cluster_charlson}) cluster and
        (\subref{fig:intervention_charlson}) intervention.
    }\label{fig:charlson}
\end{figure}

\begin{figure}
    \centering
    \begin{subfigure}{.5\imgwidth}
        \includegraphics[width=\linewidth]{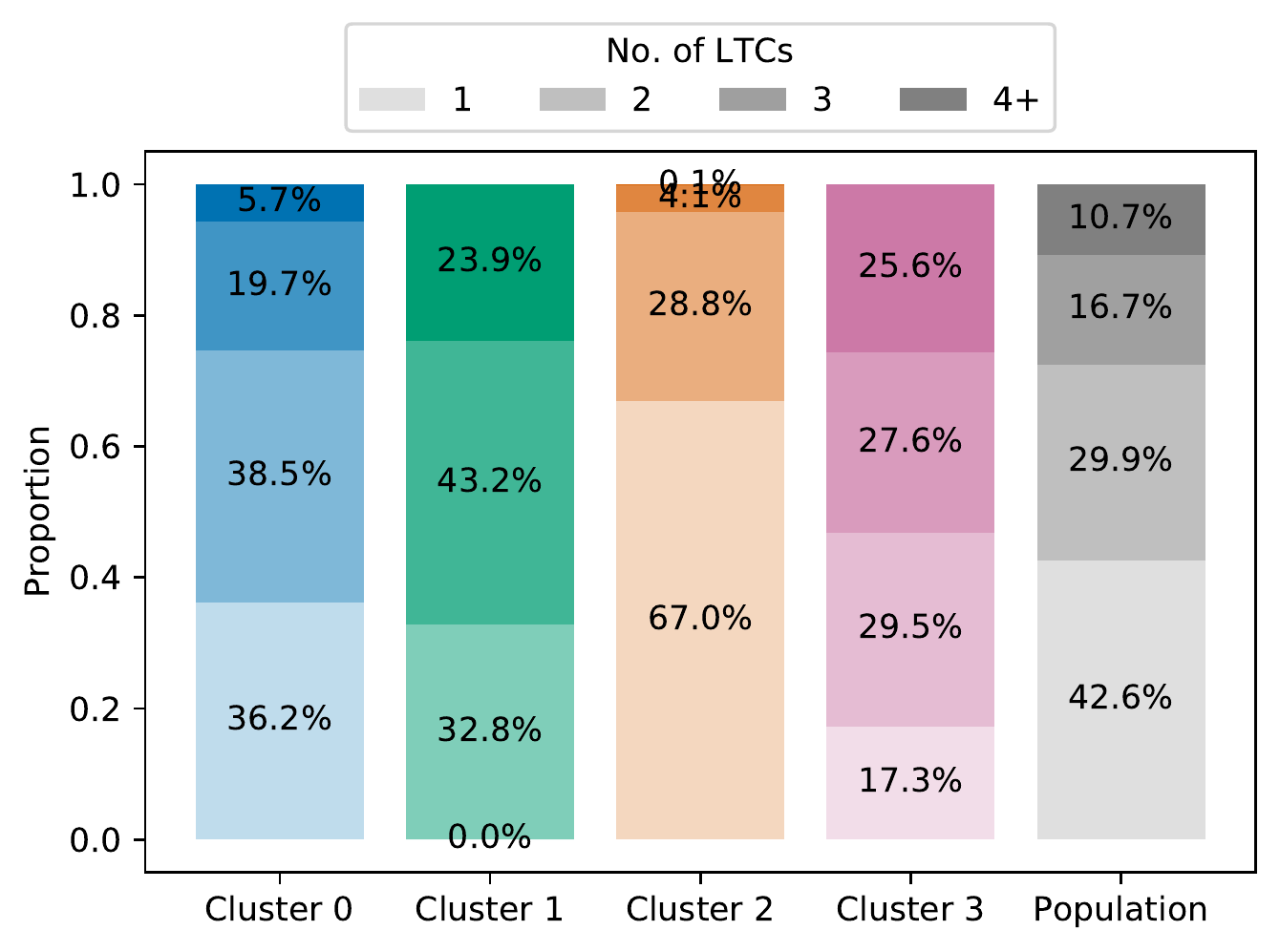}
        \caption{}\label{fig:cluster_ltcs}
    \end{subfigure}\hfill%
    \begin{subfigure}{.5\imgwidth}
        \includegraphics[width=\linewidth]{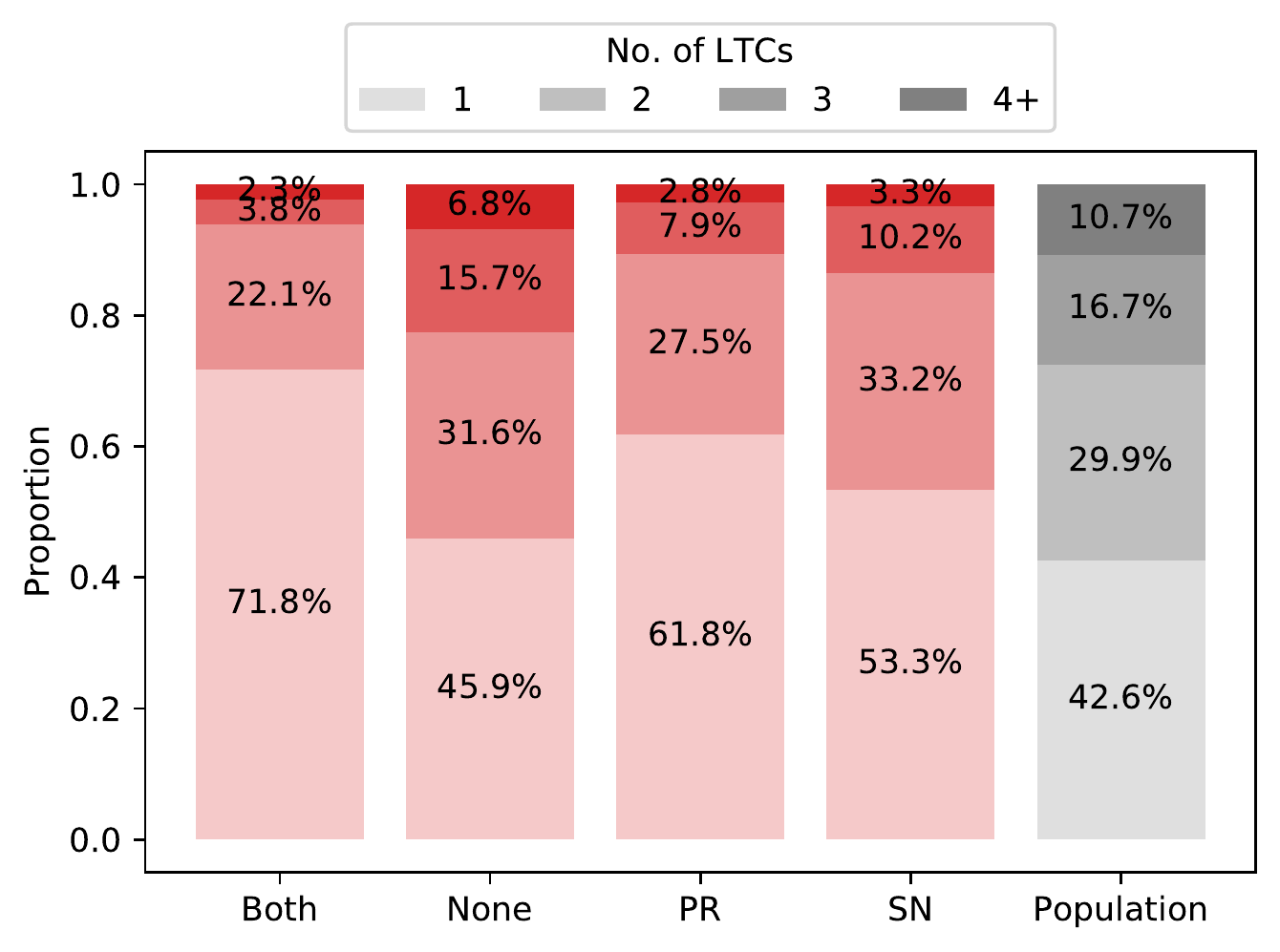}
        \caption{}\label{fig:intervention_ltcs}
    \end{subfigure}
    \caption{%
        Proportions of the number of concurrent LTCs in a spell by
        (\subref{fig:cluster_ltcs}) cluster and (\subref{fig:intervention_ltcs})
        intervention.
    }\label{fig:ltcs}
\end{figure}

\begin{figure}
    \centering
    \begin{subfigure}{.5\imgwidth}
        \includegraphics[width=\linewidth]{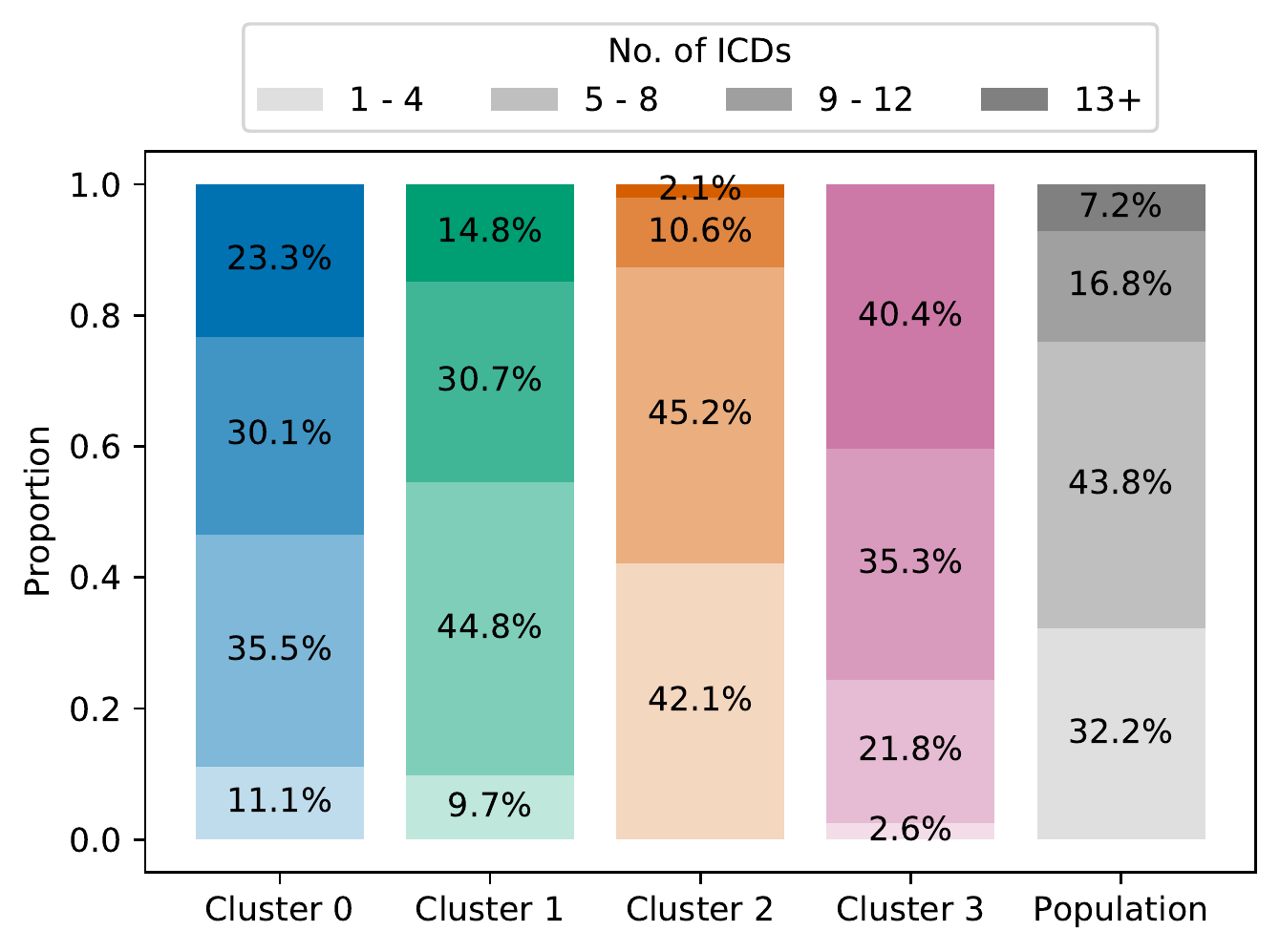}
        \caption{}\label{fig:cluster_icds}
    \end{subfigure}\hfill%
    \begin{subfigure}{.5\imgwidth}
        \includegraphics[width=\linewidth]{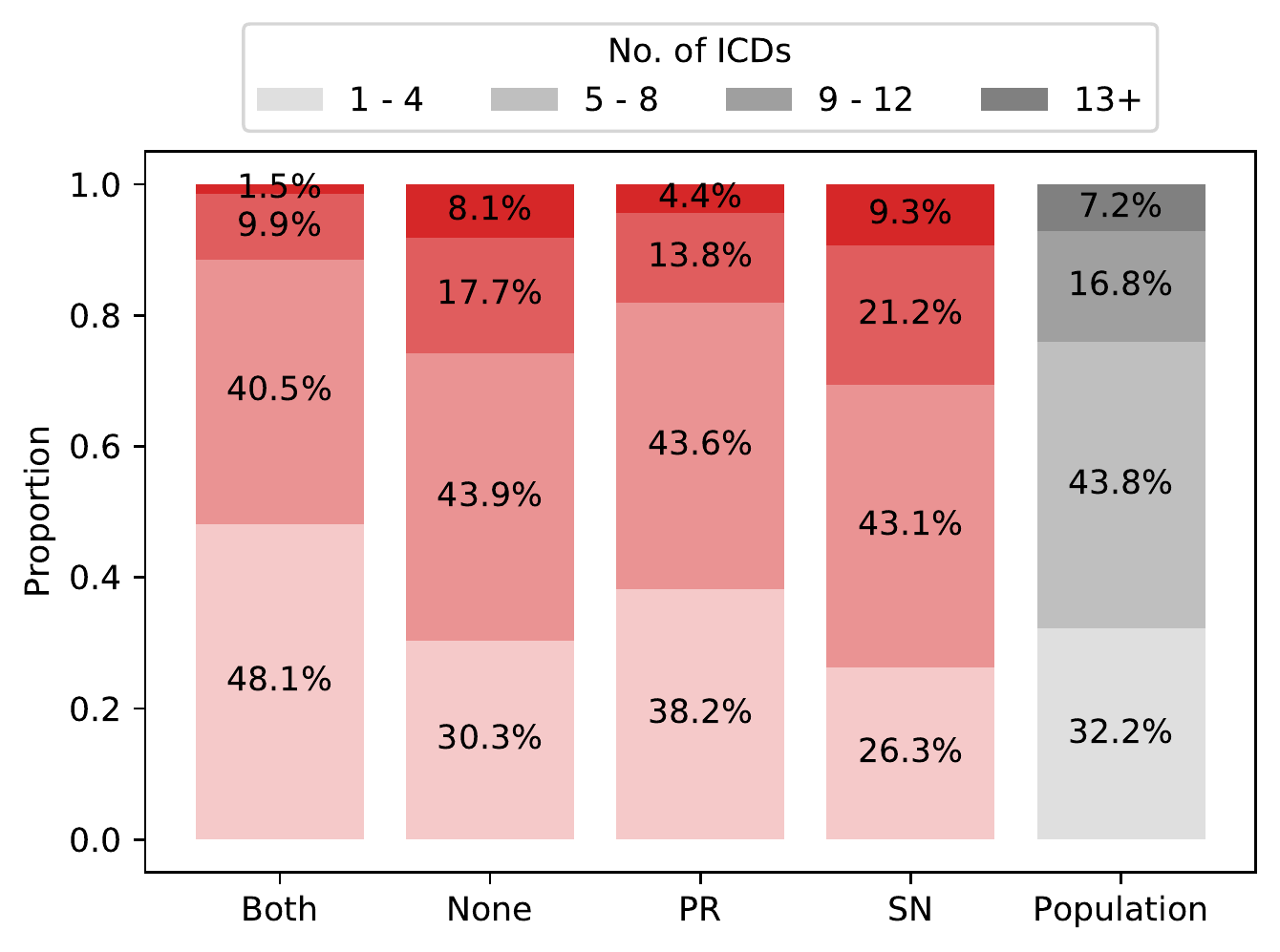}
        \caption{}\label{fig:intervention_icds}
    \end{subfigure}
    \caption{%
        Proportions of the number of concurrent ICDs in a spell by
        (\subref{fig:cluster_icds}) cluster and (\subref{fig:intervention_icds})
        intervention.
    }\label{fig:icds}
\end{figure}

The attributes listed in Table~\ref{tab:summary} can be studied beyond summaries
such as these, however. Figures~\ref{fig:los}~through~\ref{fig:icds} show the
distributions for some clinical characteristics for each cluster. Each of these
figures also shows the distribution of the same attributes when splitting the
population by intervention. While this classical approach --- of splitting a
population based on a condition or treatment --- can provide some insight into
how the different interventions are used, it has been included to highlight the
value added by segmenting the population via data without such a prescriptive
framework.

Figure~\ref{fig:los} shows the length of stay distributions as histograms.
Figure~\ref{fig:cluster_los} demonstrates the different bed resource
requirements well for each cluster --- better than Table~\ref{tab:summary} might
--- in that the difference between the clusters is not just a matter of
varying means and ranges, but entirely different shapes to their respective
distributions. Indeed, they are all positively skewed, but there is no real
consistency beyond that. When comparing this to
Figure~\ref{fig:intervention_los}, there is undoubtedly some variety, but the
overall shapes of the distributions are generally similar. The exception is the
spells with no COPD intervention where binning could not improve the
visualisation due to the widespread distribution of their lengths of stay.

The same conclusions can be drawn about spell costs from Figure~\ref{fig:cost};
there are distinct patterns between the clusters in terms of their costs, and
they align with the patterns seen in Figure~\ref{fig:los}. Such patterns are
expected given that length of stay is a driving force of healthcare costs.
Equally, there does not appear to be any immediately discernible difference in
the distribution of costs when splitting by intervention.

Similarly to the previous figures, Figure~\ref{fig:charlson} shows that
clustering has revealed distinct patterns in the CCI of the spells within each
cluster, whereas splitting by intervention does not. All clusters other than
Cluster 2 show clear, heavy tails, and in the cases of Clusters 1 and 3, the
body of the data exists far from the origin as indicated in
Table~\ref{tab:summary}. In contrast, the plots in
Figure~\ref{fig:intervention_charlson} all display similar, highly skewed
distributions regardless of intervention.

Figures~\ref{fig:ltcs}~and~\ref{fig:icds} show the proportions of each grouping
presenting levels of concurrent LTCs and ICDs, respectively. By exposing the
distribution of these attributes, some notion of the clinical complexity for
each cluster can be captured better than with Table~\ref{tab:summary} alone. In
Figure~\ref{fig:cluster_ltcs}, for instance, there are distinct LTC count
profiles among the clusters: Cluster 0 is typical of the population; Cluster 1
shows that no patient presented COPD solely as an LTC in their spells, and more
than half presented at least three; Cluster 2 is similar in form to the
population but is severely biased towards patients presenting COPD as the only
LTC; Cluster 3 is the most uniformly spread among the four bins despite the
increased length of stay and CCI suggesting a diverse array of patients in
terms of their long term medical needs.

Figure~\ref{fig:cluster_icds} largely mirrors these cluster profiles with the
number of concurrent ICDs. Some points of interest, however, are that Cluster 1
has a relatively low-leaning distribution of ICDs that does not marry up with
the high rates of LTCs, and that the vast majority of spells in Cluster 3
present with at least nine ICDs suggesting a likely wide range of conditions and
comorbidities beyond the LTCs used to calculate CCI.\

However, little can be drawn from the intervention counterparts to these figures
(i.e.\ Figures~\ref{fig:intervention_ltcs}~and~\ref{fig:intervention_icds}),
regarding the corresponding spells. One thing of note is that patients receiving
both interventions for their COPD (or either, in fact) have disproportionately
fewer LTCs and concurrent ICDs when compared to the population. Aside from this,
the profiles of each intervention are similar to one another.

As discussed earlier, the purpose of this work is to construct a queuing model
for the data described here. Insights have already been gained into the needs of
the segments that have been identified in this section. However, to glean
further insights, some parameters of the queuing model must be recovered from
the data.

\section{Constructing the queuing model}\label{sec:model}

The scarcity of data limits the options for the queuing model. However, there is
a precedent for simplifying healthcare systems to a single node with parallel
servers that emulate resource
availability.~\cite{Steins2013}~and~\cite{Williams2015} provide examples of
how this approach, when paired with discrete event simulation, can expose the
resource needs of a system beyond deterministic queuing theory models. In
particular,~\cite{Williams2015} shows how a single node, multiple server queue
can be used to accurately predict bed capacity and length of stay distributions
in a critical care unit using administrative data.

In order to follow in the suit of recent literature, this work employs a single
node using the \(M/M/c\) queue to model a hypothetical ward of patients
presenting COPD.\ In addition to this, the grouping found in
Section~\ref{subsec:overview} provides a set of patient classes in the queue.
Under this model, the following assumptions are made:
\begin{enumerate}
    \item Inter-arrival and service times of patients are each exponentially
        distributed with some mean. This distribution is used despite the system
        time distributions shown in Figure~\ref{fig:cluster_los} in order to
        simplify the model parameterisation.
    \item There are \(c \in \mathbb{N}\) servers available to arriving patients
        at the node representing the overall resource availability, including
        bed capacity and hospital staff.
    \item There is no queue or system capacity. In~\cite{Williams2015}, a
        queue capacity of zero is set under the assumption that any surplus
        arrivals would be sent to another suitable ward or unit. As this
        hypothetical ward represents COPD patients potentially throughout a
        hospital, this assumption is not held.
    \item Without the availability of expert clinical knowledge, a
        first-in-first-out service policy is employed in place of some patient
        priority framework.
\end{enumerate}

Each group of patients has its arrival distribution, the parameter of which is
the reciprocal of the mean inter-arrival times for that group. This parameter
is denoted by \(\lambda_i\) for each cluster \(i\).

Like arrivals, each group of patients has its service time distribution.
Without full details of the process order or idle periods during a spell, some
assumption must be made about the actual `service' time of a patient in the
hospital. It is assumed here that the mean service time of a group of patients
may be approximated via their mean length of stay, i.e.\ the mean time spent in
the system. For simplicity, this work assumes that for each cluster, \(i\), the
mean service time of that cluster, \(\frac{1}{\mu_i}\), is directly proportional
to the mean total system time of that cluster, \(\frac{1}{\phi_i}\), such that:
\begin{equation}\label{eq:services}
    \mu_i = p_i \phi_i
\end{equation}

\noindent where \(p_i \in \interval[open left]{0}{1}\) is some parameter to be
determined for each group.

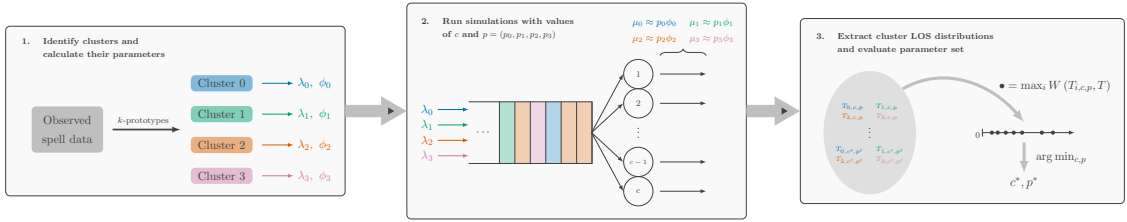
\begin{figure}
    \centering%
    \resizebox{\imgwidth}{!}{%
        \begin{tikzpicture}

    \color{black!75}
    %%%%%%%%%%%%
    % Clusters %
    %%%%%%%%%%%%
    \node[%
        draw,
        fill=gray!5,
        rounded corners,
        minimum width=110mm,
        minimum height=55mm,
    ] (clustering) at (-95mm, 17mm) {};
    \node at ([xshift=-26mm, yshift=-7mm] clustering.north) {%
        \scriptsize\textbf{%
            \begin{tabular}{rl}
                1. & Identify clusters and\\
                {} & calculate their parameters
            \end{tabular}
        }
    };

    \node[fill=gray!50, minimum width=20mm, rounded corners]
        (obs) at (-130mm, 11mm) {%
            \begin{tabular}{c}
                Observed\\
                spell data
            \end{tabular}
        };
    \node[fill=blue!50, minimum width=20mm, rounded corners]
        (c0) at (-80mm, 26mm) {Cluster 0};
    \node[fill=green!50, minimum width=20mm, rounded corners]
        (c1) at (-80mm, 16mm) {Cluster 1};
    \node[fill=orange!50, minimum width=20mm, rounded corners]
        (c2) at (-80mm, 6mm) {Cluster 2};
    \node[fill=pink!50, minimum width=20mm, rounded corners]
        (c3) at (-80mm, -4mm) {Cluster 3};

    \draw[-latex, ultra thick] ([xshift=3mm] obs.east) -- ++(22mm, 0)
        node[xshift=-1mm, midway, above] {\scriptsize\(k\)-prototypes};

    % Params
    \foreach \i/\colour/\cluster in {%
        0/blue/c0, 1/green/c1, 2/orange/c2, 3/pink/c3%
    }{%
        \node (params-\i) at ([xshift=20mm] \cluster.east)
            {\color{\colour}\(\lambda_{\i}, \ \phi_{\i}\)};
        \draw[-latex, thick, \colour]
            ([xshift=3mm] \cluster.east) -- (params-\i);
    };

    %%%%%%%%%
    % Queue %
    %%%%%%%%%
    \node[%
        draw,
        fill=gray!5,
        rounded corners,
        minimum width=110mm,
        minimum height=70mm,
    ] (queuing) at (35mm, 17mm) {};
    \node at ([xshift=-25mm, yshift=-8mm] queuing.north) {%
        \scriptsize\textbf{%
            \begin{tabular}{rl}
                2. & Run simulations with values\\
                {} & of \(c\) and \(p = \left(p_0, p_1, p_2, p_3\right)\)
            \end{tabular}
        }
    };

    \fill[orange!30] (30mm, 0) rectangle (40mm, 20mm);
    \fill[blue!30] (25mm, 0) rectangle (30mm, 20mm);
    \fill[pink!30] (20mm, 0) rectangle (25mm, 20mm);
    \fill[orange!30] (15mm, 0) rectangle (20mm, 20mm);
    \fill[green!30] (10mm, 0) rectangle (15mm, 20mm);

    \path (0, 0) pic {queue=6};
    \node (queue-in) at (0, 10mm) {};
    \node (queue-out) at (60mm, 10mm) {};

    % Arrivals
    \foreach \i/\colour in {0/blue, 1/green, 2/orange, 3/pink}{%
        \draw[-latex, \colour, thick]
            (-10mm, 17.5mm - \i * 5mm)
            to node[left, pos=0] {\color{\colour}\(\lambda_{\i}\)}
            ++(10mm, 0);
    };

    % Services
    \foreach \val in {0, 1, 3, 4}{%
        \draw[-latex, thick] (62mm, 29mm - \val * 9.5mm) -- ++(15mm, 0);
    };
    \draw[decorate, decoration={brace, amplitude=2mm}]
        (62mm, 35mm) -- ++(15mm, 0) node[midway, above=2mm] {%
            \footnotesize%
            \begin{tabular}{cc}
                \color{blue}{\(\mu_0 \approx p_0\phi_0\)} &
                \color{green}{\(\mu_1 \approx p_1\phi_1\)}\\
                \color{orange}{\(\mu_2 \approx p_2\phi_2\)} &
                \color{pink}{\(\mu_3 \approx p_3\phi_3\)}\\
            \end{tabular}
        };

    %%%%%%%%%%
    % Output %
    %%%%%%%%%%
    \node[%
        draw,
        fill=gray!5,
        rounded corners,
        minimum width=105mm,
        minimum height=60mm,
    ] (output) at (160mm, 17mm) {};
    \node at ([xshift=-18mm, yshift=-8mm] output.north) {%
        \scriptsize\textbf{%
            \begin{tabular}{rl}
                3. & Extract cluster LOS distributions\\
                {} & and evaluate parameter set
            \end{tabular}
        }
    };

    \node[ellipse, fill=gray!25, minimum width=30mm, minimum height=40mm]
        (times) at (130mm, 10mm) {};
    \node (top) at ([yshift=7mm] times) {%
        \tiny\begin{tabular}{cc}
            \color{blue}\(T_{0,c,p}\) &
            \color{green}\(T_{1,c,p}\)\\
            \color{orange}\(T_{2,c,p}\) &
            \color{pink}\(T_{3,c,p}\)
        \end{tabular}
    };
    \node at ([yshift=1mm] times) {\(\vdots\)};
    \node (bottom) at ([yshift=-7mm] times) {%
        \tiny\begin{tabular}{cc}
            \color{blue}\(T_{0,c',p'}\) &
            \color{green}\(T_{1,c',p'}\)\\
            \color{orange}\(T_{2,c',p'}\) &
            \color{pink}\(T_{3,c',p'}\)
        \end{tabular}
    };

    \node[label={[below] \scriptsize\(0\)}]
        (zero) at (165mm, 10mm) {};
    \draw[|-latex, thick] (zero.east) -- ++(30mm, 0);
    \foreach \val in {3.2mm, 5.1mm, 7.7mm, 10.3mm, 13mm, 19.4mm, 23mm}{%
        \draw[fill] ([xshift=\val] zero.east) circle (.5mm);
    };

    \draw[mypointer]
        (times.north east) to [out=30, in=130]
        ([xshift=15mm, yshift=2mm] zero.north)
        node[xshift=10mm, above=8mm] {%
            \color{black!75}\(\bullet = \max_{i} W\left(T_{i,c,p}, T\right)\)
        };

    \node (params) at ([xshift=15mm, yshift=-15mm] zero.south) {%
        \large\(c^*, p^*\)
    };

    \draw[mypointer]
        ([yshift=10mm] params.north) -- (params.north)
        node[xshift=12mm, yshift=5mm] {\color{black!75}\(\argmin_{c,p}\)};

    %%%%%%%%%%
    % Arrows %
    %%%%%%%%%%
    \draw[myarrow] (clustering.east) -- (queuing.west);
    \draw[myarrow] (queuing.east) -- (output.west);

    \end{tikzpicture}
    }
    \caption{%
        A diagrammatic depiction of the queuing parameter recovery process.
    }\label{fig:process}
\end{figure}

One of the few ground truths available in the provided data is the distribution
of the total length of stay. Given that the length of stay and resource
availability are connected, the approach here will be to simulate the length of
stay distribution for a range of values \(p_i\) and \(c\), to find the
parameters that best match the observed data. Figure~\ref{fig:process} provides
a diagrammatic depiction of the process described in this section.

Several methods are available for the statistical comparison of two or more
distributions, such as the Kolmogorov-Smirnov test, a variety of discrepancy
approaches such as summed mean-squared error, and \(f\)-divergences. A popular
choice among the last group (which may be considered distance-like) is the
Kullback-Leibler divergence which measures relative information entropy from one
probability distribution to another~\cite{Kullback1951}. A key issue with many
of these methods is that they lack interpretability, something which is
paramount when conveying information to stakeholders, not just from explaining
how something works but also how its results may be explained.

As such, a reasonable candidate is the (first) Wasserstein metric, also known as
the `earth mover' or `digger' distance~\cite{Vaserstein1969}. The Wasserstein
metric satisfies the conditions of a formal mathematical metric (like the
typical Euclidean distance), and its values take the units of the distributions
under comparison (in this case: days). These characteristics can aid
understanding and explanation. In simple terms, the distance measures the
approximate `minimal work' required to move between two probability
distributions where `work' can be loosely defined as the product of how much of
the distribution's mass moves and the distance by which it must be moved. More
formally, the Wasserstein distance between two probability distributions \(U\)
and \(V\) is defined as:
\begin{equation}\label{eq:wasserstein}
    W(U, V) = \int_{0}^{1} \left\vert F^{-1}(t) - G^{-1}(t) \right\vert dt
\end{equation}

\noindent where \(F\) and \(G\) are the cumulative density functions of \(U\)
and \(V\), respectively. A proof of~\eqref{eq:wasserstein} is presented
in~\cite{Ramdas2017}. The parameter set with the smallest maximum distance
between any cluster's simulated system time distribution and the overall
observed length of stay distribution is then taken to be the most appropriate.
To be specific, let \(T\) denote the system time distribution of all of the
observed data and let \(T_{i,c,p}\) denote the system time distribution for
cluster \(i\) obtained from a simulation with \(c\) servers and \(p :=
\left(p_0, p_1, p_2, p_3\right)\). Then the optimal parameter set \(\left(c^*,
p^*\right)\) is given by:
\begin{equation}\label{eq:parameters}
    \left(c^*, p^*\right) = \argmin_{c, p} \left\{%
        \max_{i} \left\{ W\left(T_{i,c,p}, T\right) \right\}%
    \right\}
\end{equation}

The parameter sweep included values of each \(p_i\) from \(0.5\) to \(1.0\) with
a granularity of \(5.0 \times 10^{-2}\) and values of \(c\) from \(40\) to
\(60\) at steps of five. These choices were informed by the assumptions of the
model and formative analysis to reduce the parameter space given the
computational resources required to conduct the simulations. Each parameter set
was repeated 50 times with each simulation running for four years of virtual
time. The warm-up and cool-down periods were taken to be approximately one year
each leaving two years of simulated data from each repetition.

\begin{figure}
    \centering%
    \begin{subfigure}{.5\imgwidth}
        \includegraphics[width=\linewidth]{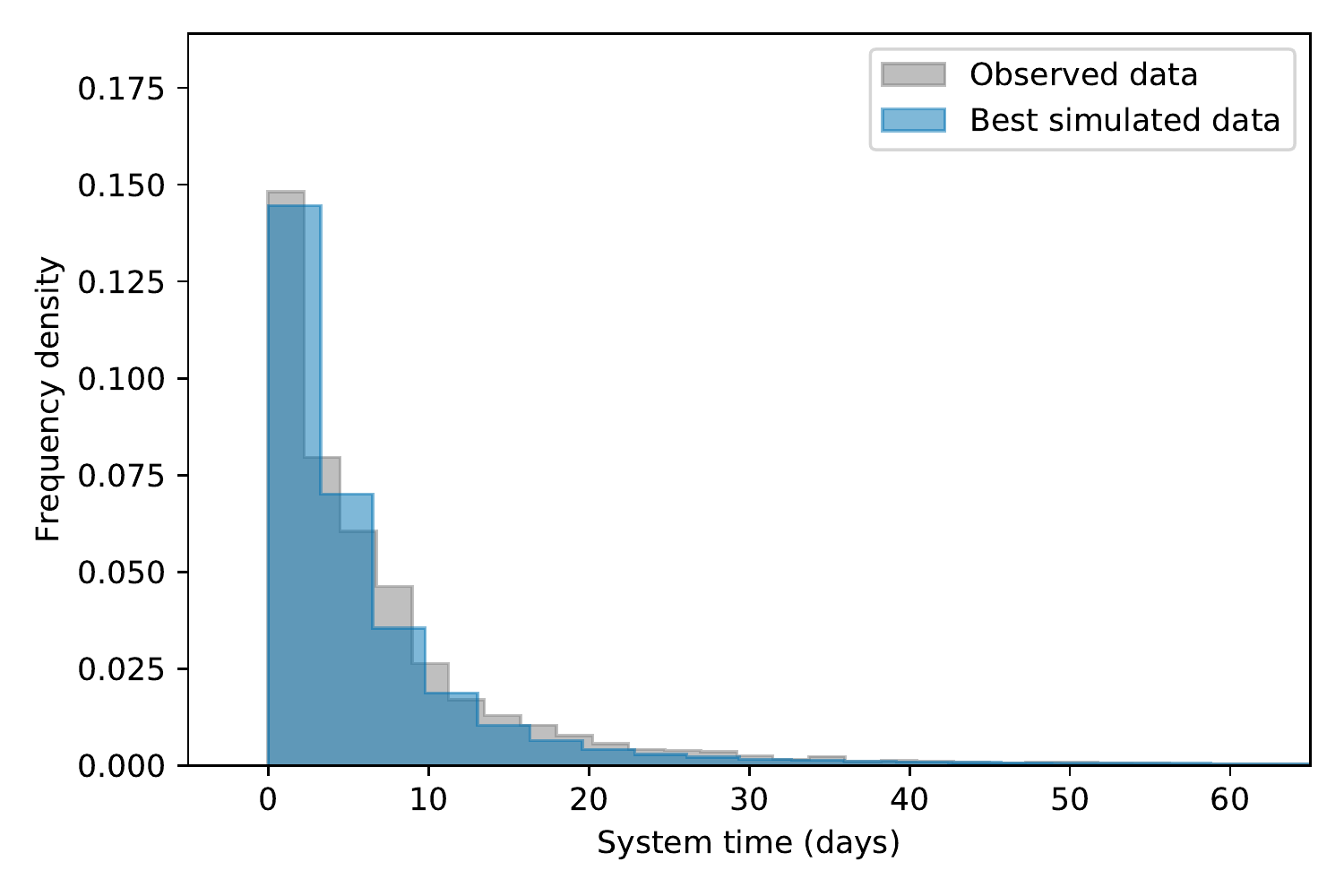}
        \caption{}\label{fig:best_params}
    \end{subfigure}\hfill%
    \begin{subfigure}{.5\imgwidth}
        \includegraphics[width=\linewidth]{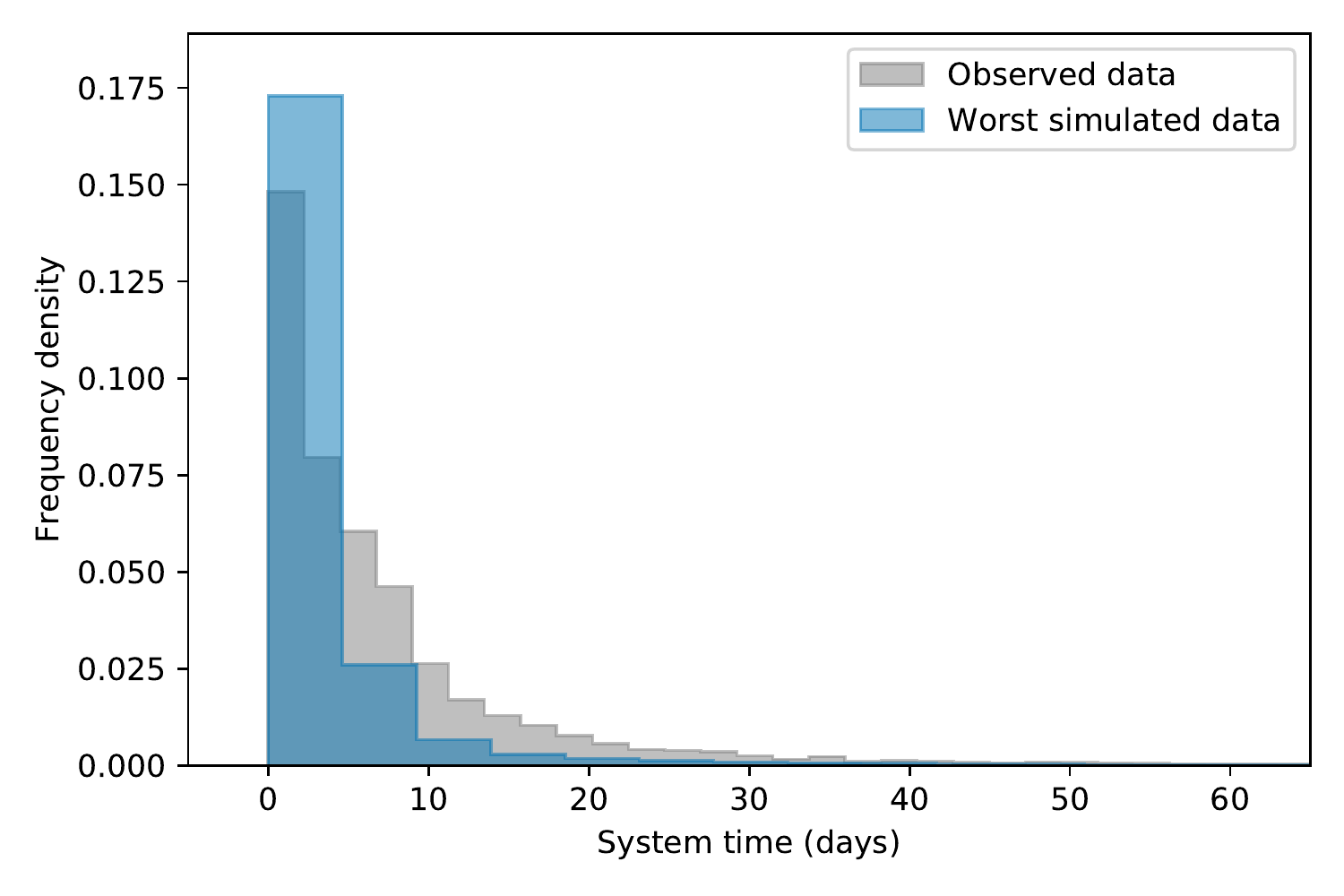}
        \caption{}\label{fig:worst_params}
    \end{subfigure}
    \caption{Histograms of the simulated and observed length of stay data for
             the (\subref{fig:best_params}) best and (\subref{fig:worst_params})
             worst parameter sets.}\label{fig:params}
\end{figure}

The results of this parameter sweep can be summarised in
Figure~\ref{fig:params}. Each plot shows a comparison of the observed lengths of
stay across all groups and the newly simulated data with the best and worst
parameter sets, respectively. In the best case, a very close fit has been found.
Meanwhile, Figure~\ref{fig:worst_params} highlights the importance of good
parameter estimation under this model since the likelihood of short-stay patient
arrivals has been inflated disproportionately against the tail of the
distribution. Table~\ref{tab:comparison} reinforces these results numerically,
showing a precise fit by the best parameters across the board.

\begin{table}
    \centering
    \resizebox{\tabwidth}{!}{%
        \begin{tabular}{lrrrrrrrrrrrrr}
        \toprule
        {} & \multicolumn{6}{l}{Model parameter and result} & \multicolumn{7}{l}{LOS statistic} \\
        {} &                    \(p_0\) & \(p_1\) & \(p_2\) & \(p_3\) & \(c\) & Max. distance &          Mean &   Std. &  Min. &   25\% &  Med. &   75\% &    Max. \\
        \midrule
        Observed        &                        NaN &     NaN &     NaN &     NaN &   NaN &          0.00 &          7.70 &  11.86 & -0.02 &  1.49 &  4.20 &  8.93 &  224.93 \\
        Best simulated  &                       0.95 &     1.0 &     1.0 &     0.5 &  40.0 &          1.28 &          7.00 &  12.09 &  0.00 &  1.44 &  3.57 &  7.65 &  326.46 \\
        Worst simulated &                       0.50 &     0.5 &     0.5 &     1.0 &  40.0 &          4.25 &          4.36 &  13.40 &  0.00 &  0.72 &  1.78 &  3.84 &  463.01 \\
        \bottomrule
        \end{tabular}
    }
    \caption{A comparison of the observed data, and the best and worst simulated
        data based on the model parameters and summary statistics for length of
    stay (LOS).}\label{tab:comparison}
\end{table}

In this section, the previously identified clustering enriched the overall
queuing model and was used to recover the parameters for several classes within
that. Now, using this model, the next section details an investigation into the
underlying system by adjusting the parameters of the queue with the clustering.

\section{Adjusting the queuing model}\label{sec:scenarios}

This section comprises several what-if scenarios --- a classic component of
healthcare operational research --- under the novel parameterisation of the
queue established in Section~\ref{sec:model}. The outcomes of interest in this
work are server (resource) utilisation and system times. These metrics capture
the driving forces of cost and the state of the system. Specifically, the
objective of these experiments is to address the following questions:
\begin{itemize}
    \item How would the system be affected by a change in overall patient
        arrivals?
    \item How is the system affected by a change in resource availability (i.e.\
        a change in \(c\))?
    \item How is the system affected by patients moving between clusters?
\end{itemize}

Given the nature of the observed data, the queuing model parameterisation and
its assumptions, the effects on the chosen metrics in each scenario are in
relative terms with respect to the base case. The base case being those results
generated from the best parameter set recorded in Table~\ref{tab:comparison}. In
particular, the data from each scenario is scaled by the corresponding median
value in the base case, meaning that a metric having a value of 1 is ‘normal’.

As mentioned in Section~\ref{sec:intro}, the source code used throughout this
work is available has been archived online~\cite{Wilde2020github}. Also, the
datasets generated from the simulations in this section, and the parameter
sweep, have been archived online~\cite{Wilde2020results}.

\subsection{Changes to overall patient arrivals}\label{subsec:arrivals}

Changes in overall patient arrivals to a queue reflect real-world scenarios
where some stimulus is improving (or worsening) the condition of the patient
population. Examples of stimuli could include an ageing population or
independent life events that lead to a change in deprivation, such as an
accident or job loss. Within this model, overall patient arrivals are altered
using a scaling factor denoted by \(\sigma\in\mathbb{R}\). This scaling factor
is applied to the model by multiplying each cluster's arrival rate by
\(\sigma\). That is, for cluster \(i\), its new arrival rate, \(\hat\lambda_i\),
is given by:
\begin{equation}\label{eq:lambda}
    \hat\lambda_{i} = \sigma\lambda_i
\end{equation}

\begin{figure}
    \centering
    \begin{subfigure}{.5\imgwidth}
        \includegraphics[width=\linewidth]{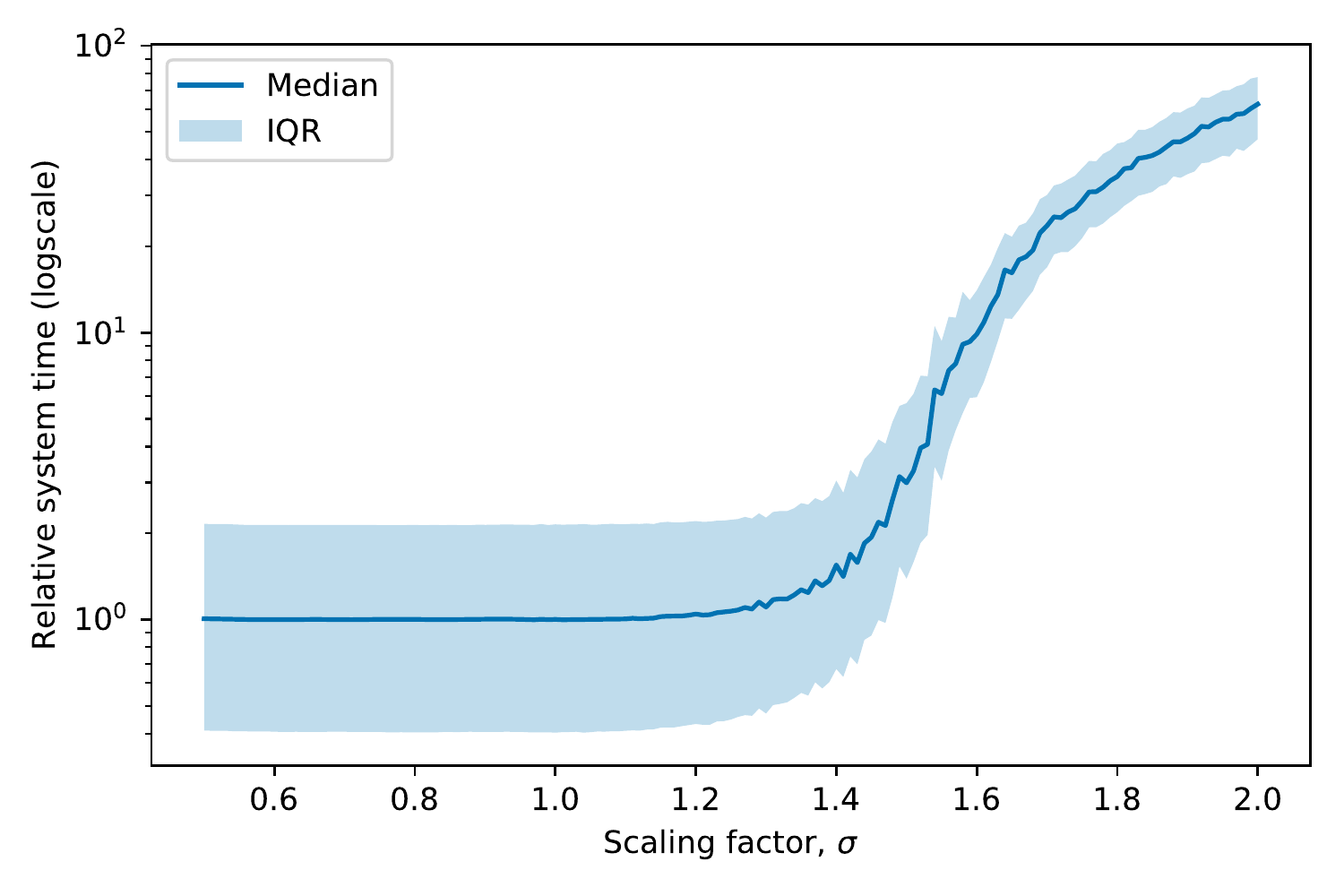}
        \caption{}\label{fig:lambda_time}
    \end{subfigure}\hfill%
    \begin{subfigure}{.5\imgwidth}
        \includegraphics[width=\linewidth]{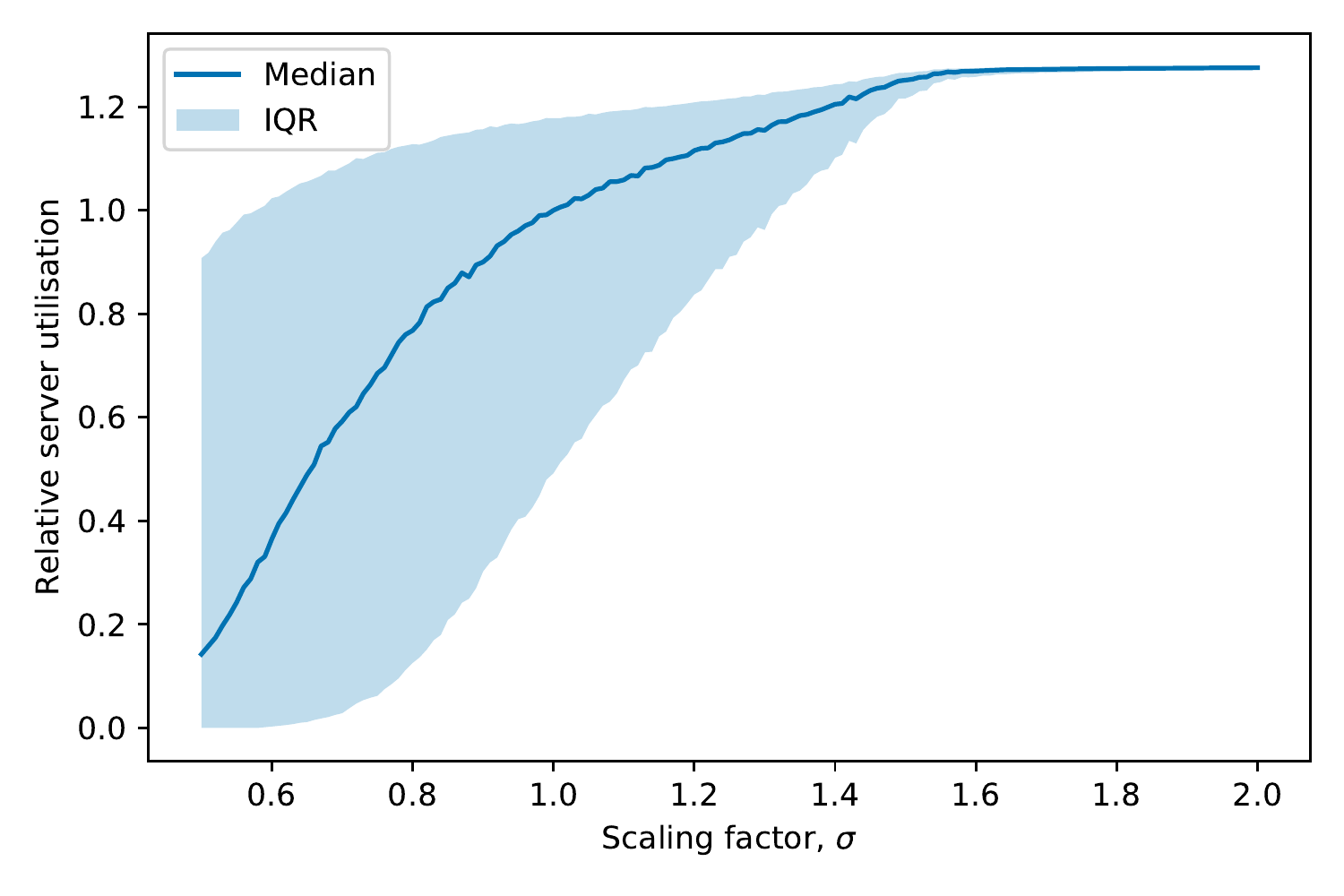}
        \caption{}\label{fig:lambda_util}
    \end{subfigure}
    \caption{%
        Plots of \(\sigma\) against relative (\subref{fig:lambda_time})~system
        time and (\subref{fig:lambda_util})~server utilisation.
    }\label{fig:lambda}
\end{figure}

Figure~\ref{fig:lambda} shows the effects of changing patient arrivals on
(\subref{fig:lambda_time})~relative system times and
(\subref{fig:lambda_util})~relative server utilisation for values of \(\sigma\)
from 0.5 to 2.0 at a precision of \(1.0 \times 10^{-2}\). Specifically, each
plot in the figure (and the subsequent figures in this section) shows the median
and interquartile range (IQR) of each relative attribute. These metrics provide
an insight into the experience of the average user (or server) in the system.
Furthermore, they reveal the stability or variation of the body of users
(servers).

What is evident from these plots is that things are happening as one might
expect: as arrivals increase, the strain on the system increases. However, it
should be noted that it also appears that the model has some amount of slack
relative to the base case. Looking at Figure~\ref{fig:lambda_time}, for
instance, the relative system times (i.e.\ the relative length of stay for
patients) remains unchanged up to \(\sigma \approx 1.2\), or an approximate 20\%
increase in arrivals of COPD patients. Beyond that, relative system times rise
to an untenable point where the median time becomes orders of magnitude above
the norm.

However, Figure~\ref{fig:lambda_util} shows that the situation for the system's
resources reaches its worst-case near to the start of that spike in relative
system times (at \(\sigma \approx 1.4\)). That is, the median server utilisation
reaches a maximum (this corresponds to constant utilisation) at this point, and
the variation in server utilisation disappears entirely.

\subsection{Changes to resource availability}\label{subsec:resources}

As is discussed in Section~\ref{sec:model}, the resource availability of the
system is captured by the number of parallel servers, \(c\). Therefore, to
modify the overall resource availability, only the number of servers needs to be
changed. This kind of sensitivity analysis is usually done to determine the
opportunity cost of adding service capacity to a system, e.g.\ would an increase
of \(n\) servers increase efficiency without exceeding a budget?

To reiterate the beginning of this section: all suitable parameters are given in
relative terms, including the number of servers here. By doing this, the
changes in resource availability are more easily seen, and do away with any
concerns as to what a particular number of servers precisely reflects in the
real world.

\begin{figure}
    \centering
    \begin{subfigure}{.5\imgwidth}
        \includegraphics[width=\linewidth]{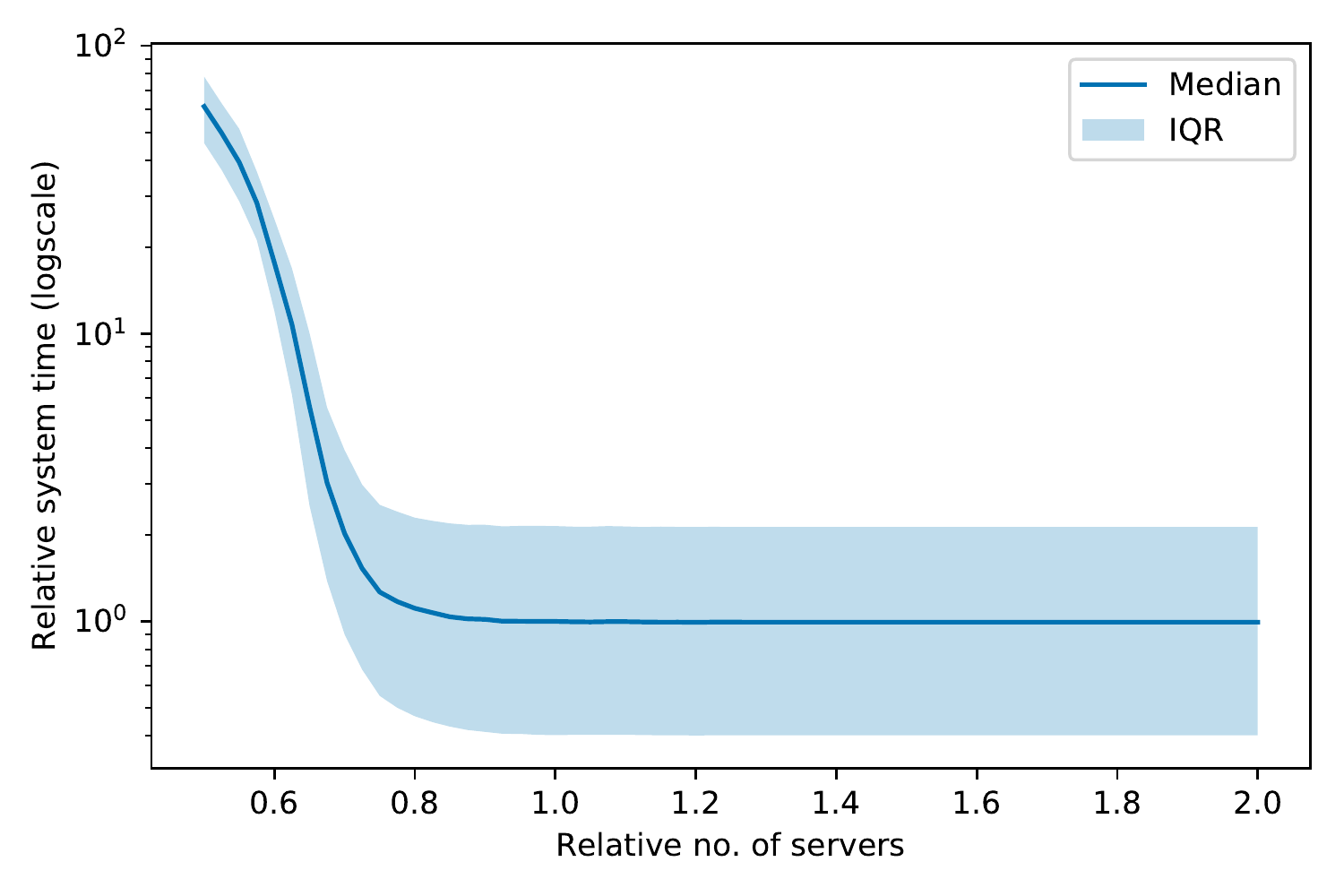}
        \caption{}\label{fig:servers_time}
    \end{subfigure}\hfill%
    \begin{subfigure}{.5\imgwidth}
        \includegraphics[width=\linewidth]{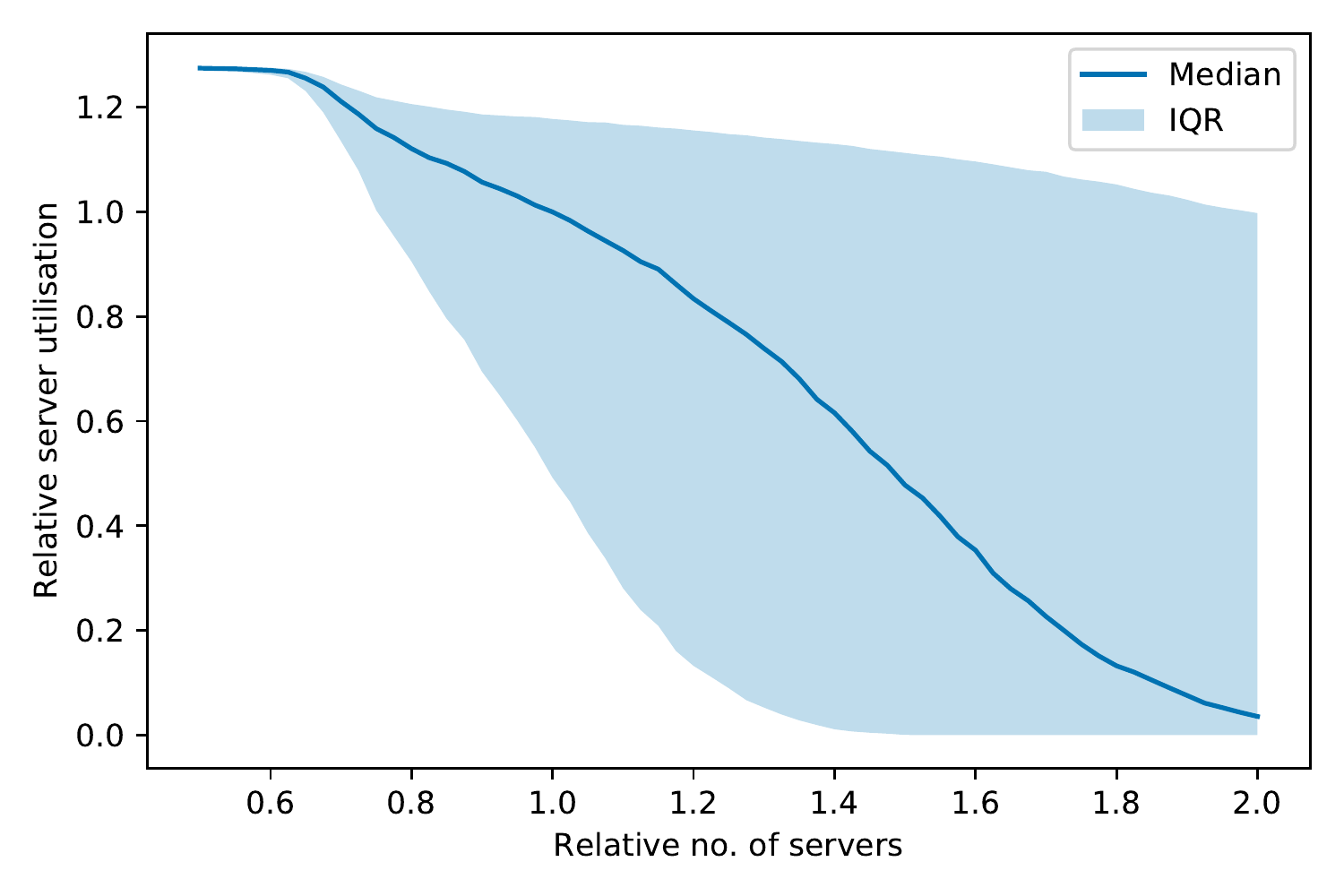}
        \caption{}\label{fig:servers_util}
    \end{subfigure}
    \caption{%
        Plots of the relative number of servers against relative
        (\subref{fig:servers_time})~system time and
        (\subref{fig:servers_util})~server utilisation.
    }\label{fig:servers}
\end{figure}

Figure~\ref{fig:servers} shows how the relative resource availability affects
relative system times and server utilisation. In this scenario, the relative
number of servers took values from 0.5 to 2.0 at steps of \(2.5 \times 10^{-2}\)
--- this is equivalent to a step size of one in the actual number of servers.
Overall, these figures fortify the claim from the previous scenario that there
is some room to manoeuvre so that the system runs `as normal' but pressing on
those boundaries results in massive changes to both resource requirements and
system times.

In Figure~\ref{fig:servers_time} this amounts to a maximum of 20\% slack in
resources before relative system times are affected; further reductions quickly
result in a potentially tenfold increase in the median system time, and up to 50
times once resource availability falls by 50\%. Moreover, the variation in the
body of the relative times (i.e.\ the IQR) decreases as resource availability
decreases. The reality of this is that patients arriving at a hospital are
forced to consume more significant amounts of resources (by merely being in a
hospital) regardless of their condition, putting added strains on the system.

Meanwhile, it appears that there is no tangible change in relative system times
given an increase in the number of servers. This indicates that the model
carries sufficient resources to cater to the population under normal
circumstancesand that adding service capacity will not necessarily improve
system times.

Again, Figure~\ref{fig:servers_util} shows that there is a substantial change in
the variation in the relative utilisation of the servers. In this case, the
variation dissipates as resource levels fall and increase as they increase.
While the relationship between real hospital resources and the number of servers
is not exact, having variation in server utilisation would suggest that parts of
the system may be configured or partitioned away in the case of some significant
public health event (such as a global pandemic) without overloading the system.

\subsection{Moving arrivals between clusters}\label{subsec:moving}

This scenario is perhaps the most relevant to actionable public health research
of those presented here. The clusters identified in this work could be
characterised by their clinical complexities and resource requirements, as done
in Section~\ref{subsec:overview}. Therefore, being able to model the movement of
some proportion of patient spells from one cluster to another will reveal how
those complexities and requirements affect the system itself. The reality is
then that if some public health policy could be implemented to enact that
movement informed by a model such as this, then real change would be seen in the
real system.

In order to model the effects of spells moving between two clusters, the
assumption is that services remain the same (and so does each cluster's
\(p_i\)), but their arrival rates are altered according to some transfer
proportion. Consider two clusters indexed at \(i, j\), and their respective
arrival rates, \(\lambda_i, \lambda_j\), and let \(\delta \in [0, 1]\) denote
the proportion of arrivals to be moved from cluster \(i\) to cluster \(j\). Then
the new arrival rates for each cluster, denoted by \(\hat\lambda_i,
\hat\lambda_j\) respectively, are:
\begin{equation}\label{eq:moving}
    \hat\lambda_i = \left(1 - \delta\right) \lambda_i
    \quad \text{and} \quad
    \hat\lambda_j = \delta\lambda_i + \lambda_j
\end{equation}

By moving patient arrivals between clusters in this way, the overall arrivals
are left the same since the sum of the arrival rates is the same. Hence, the
(relative) effect on server utilisation and system time can be measured
independently.

\begin{figure}
    \centering
    \includegraphics[width=\imgwidth]{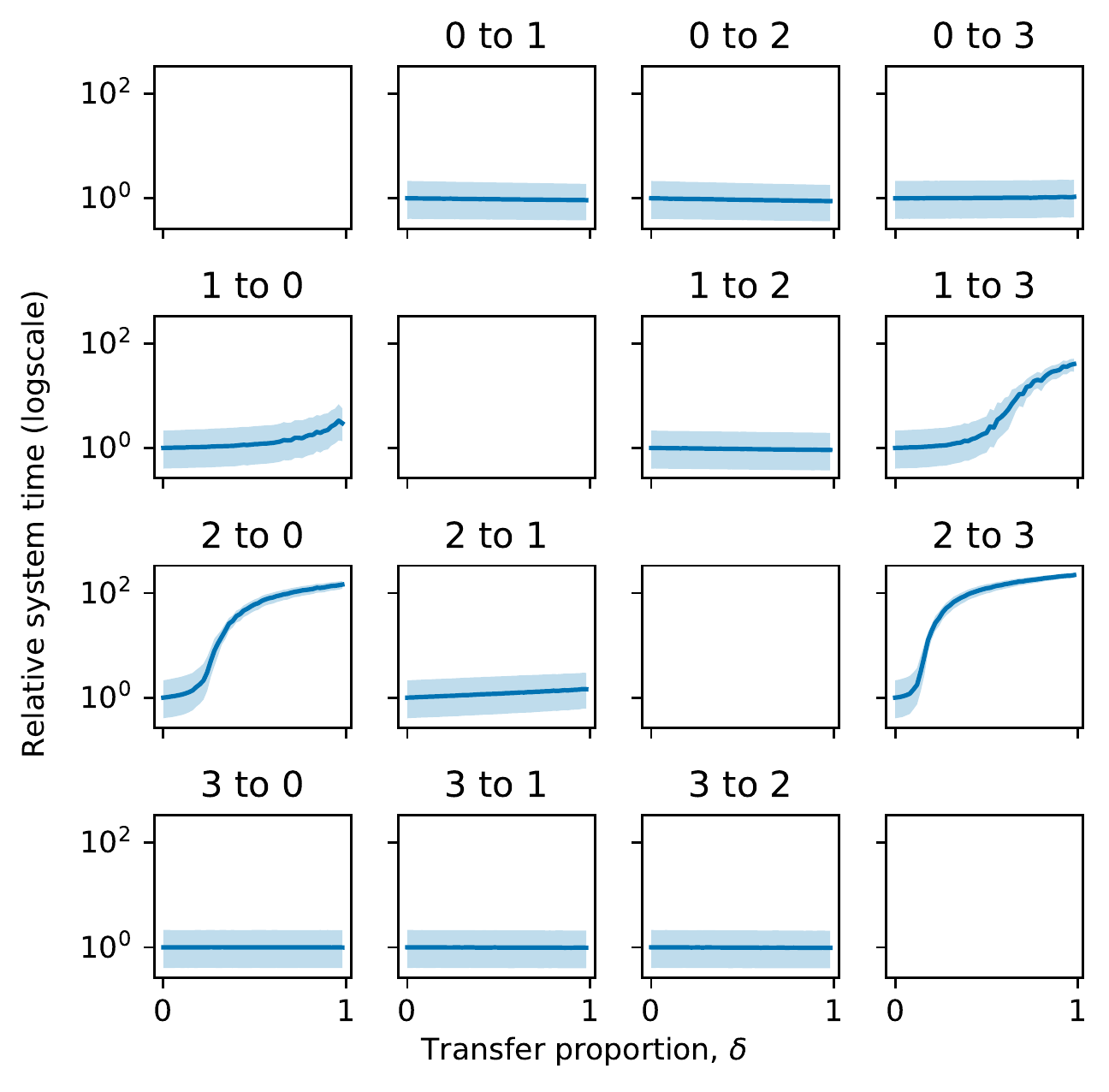}
    \caption{%
        Plots of proportions of each cluster moving to another against relative
        system time.
    }\label{fig:moving_time}
\end{figure}

\begin{figure}
    \centering
    \includegraphics[width=\imgwidth]{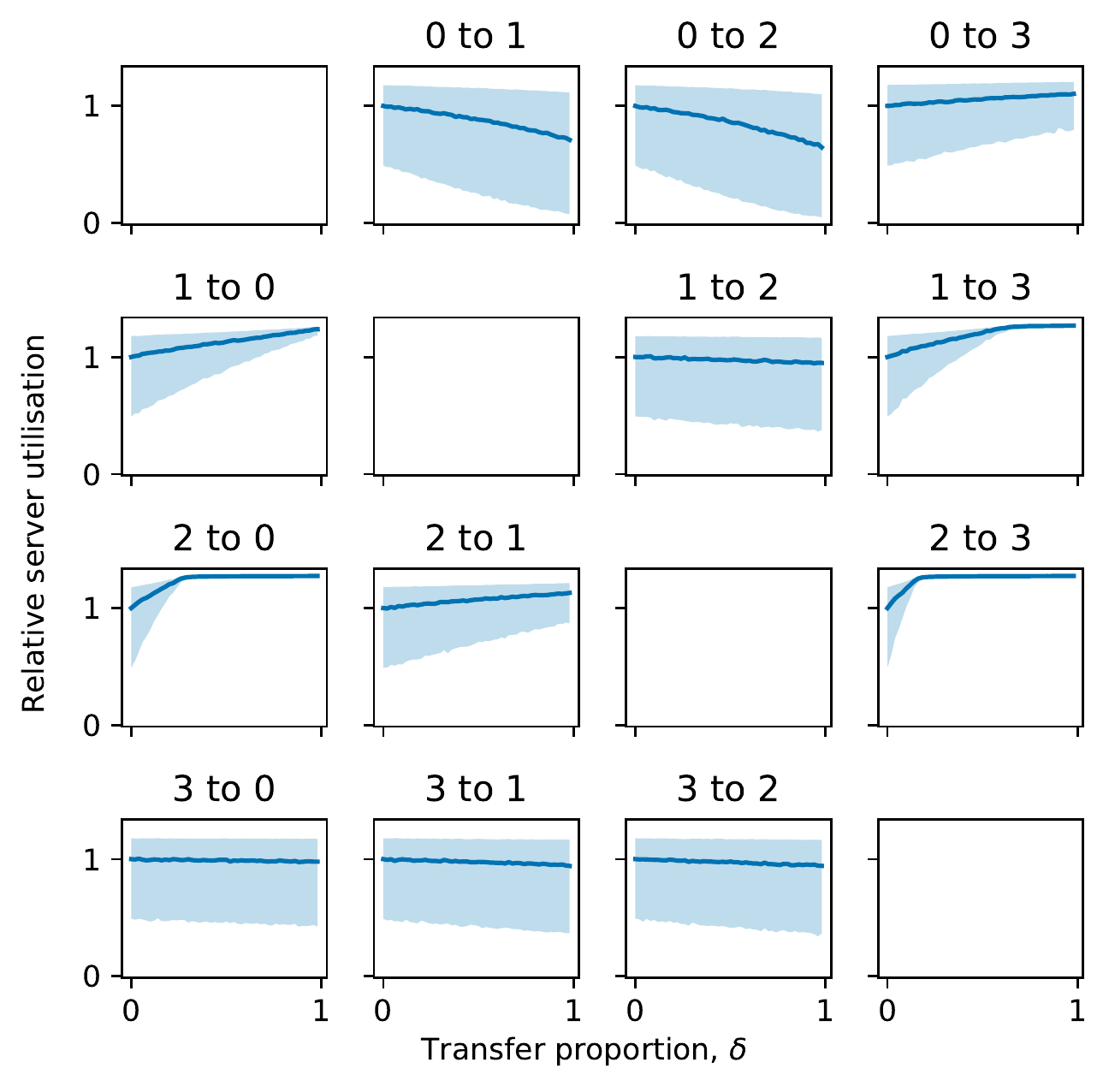}
    \caption{%
        Plots of proportions of each cluster moving to another on relative
        server utilisation.
    }\label{fig:moving_util}
\end{figure}

Figures~\ref{fig:moving_time}~and~\ref{fig:moving_util} show the effect of
moving patient arrivals between clusters on relative system time and relative
server utilisation, respectively. In each figure, the median and IQR for the
corresponding attribute is shown, as in the previous scenarios. Each scenario
was simulated using values of \(\delta\) from 0.0 to 1.0 at steps of \(2.0
\times 10^{-2}\).

Considering Figure~\ref{fig:moving_time}, it is clear that there are some cases
where reducing particular types of spells (by making them like another type of
spell) does not affect overall system times. Namely, moving the high resource
requirement spells that describe Cluster 0 and Cluster 3 to any other cluster.
These clusters make up only 10\% of all arrivals, and this figure shows that in
terms of system times, the model can handle them without concern under normal
conditions. The concern comes when either of the other clusters moves to Cluster
0 or Cluster 3. Even as few as one in five of the low complexity, low resource
needs arrivals in Cluster 2 moving to either cluster results in large jumps in
the median system time for all arrivals, and soon after, as, in the previous
scenario, any variation in the system times disappears indicating an overborne
system.

With relative server utilisation, the story is much the same. The ordinary
levels of high complexity, high resource arrivals from Cluster 3 are absorbed by
the system and moving these arrivals to another cluster bears no effect on
resource consumption levels. Likewise, either of the low-resource needs clusters
moving even slightly toward high resource requirements completely overruns the
system’s resources. However, the relative utilisation levels of the system
resources can be reduced by moving arrivals from Cluster 0 to either Cluster 1
or Cluster 2, i.e.\ by reducing the overall resource requirements of such spells. 

In essence, this entire analysis offers two messages: that there are several
ways in which the system can get worse and even overwhelmed but, more
importantly, that any meaningful impact on the system must come from a stimulus
outside of the system that results in more healthy patients arriving at the
hospital. This conclusion is non-trivial; the first two scenarios in this
analysis show that there are no quick solutions to reduce the effect of COPD
patients on hospital capacity or length of stay. The only effective intervention
is found through inter-cluster transfers.

\section{Conclusion}\label{sec:conclusion}

This work presents a novel approach to investigating a healthcare population
that encompasses the topics of segmentation analysis, queuing models, and the
recovery of queuing parameters from incomplete data. This investigation is done
despite characteristic limitations in operational research concerning the
availability of fine-grained data, and this work only uses administrative
hospital spell data from patients presenting COPD from the Cwm Taf Morgannwg
UHB.\

By considering a variety of attributes present in the data, and engineering
some, a useful clustering of the spell population is identified that
successfully feeds into a multi-class, \(M/M/c\) queue to model a hypothetical
COPD ward. With this model, several insights are gained by investigating
purposeful changes in the parameters of the model that have the potential to
inform actual public health policy. 

In particular, since neither the resource capacity of the system nor the
clinical processes of the spells are evident in the data, service times and
resource levels are not available. However, the length of stay is. Using what is
available, this work assumes that mean service times can be parameterised using
mean lengths of stay. By using the Wasserstein distance to compare the
distribution of the simulated lengths of stay data with the observed data, a
best performing parameter set is found via a parameter sweep. 

This parameterisation ultimately recovers a surrogate for service times for each
cluster, and a universal number of servers to emulate resource availability. The
parameterisation itself offers its strengths by being simple and effective.
Despite its simplicity, a good fit to the observed data is found, and --- as is
evident from the closing section of this work --- substantial and useful
insights can be gained into the needs of the population under study. 

This mode of analysis, in effect, considers all types of patient arrivals and
how they each impact the system in terms of resource capacity and length of
stay. By investigating scenarios into changes in both overall patient arrivals
and resource capacity, it is clear that there is no quick solution to be
employed from within the hospital to improve COPD patient spells. The only
effective, non-trivial intervention is to improve the overall health of the
patients arriving at the hospital, as is shown by moving patient arrivals
between clusters. In reality, this would correspond to an external, preventative
policy that improves the overall health of COPD patients. 

\section*{Acknowledgements}

The authors wish to thank the Cwm Taf Morgannwg University Health Board for
their funding and support of the PhD of which this work has formed a part.

\end{document}